\documentclass[12pt]{iopart}

\usepackage{hyperref}
\usepackage{iopams}
\usepackage{url}
\usepackage{units}
\usepackage{graphicx}
\usepackage{color}
\newcommand{\eqref}[1]{(\ref{#1})}

\newcommand{\sub}[1]{_{\mathrm{#1}}}
\newcommand{\parl}{\parallel}
\newcommand{\non}{\nonumber}

\newcommand{\val}[2]{\unit[#1]{#2}}
\newcommand{\leftm}{\left( \begin{matrix}}
\newcommand{\rightm}{\end{matrix} \right)}
\newcommand{\lw}{\linewidth}

\newcommand{\mnu}{m_{\nu_e}}
\newcommand{\mnusq}{\mnu^2}

\newcommand{\Eep}{E_0}
\newcommand{\zsource}{z\sub{source}}

\newcommand{\diff}{\mathrm{d}}
\newcommand{\Erel}{E\sub{rel}}
\newcommand{\Epot}{E\sub{pot}}
\newcommand{\Etot}{E\sub{tot}}
\newcommand{\Ekin}{E\sub{kin}}
\newcommand{\eem}{E}
\newcommand{\aem}{\theta}
\newcommand{\amax}{\aem\sub{max}}
\newcommand{\meff}{m\sub{eff}}
\newcommand{\angledist}{g(\aem)}
\newcommand{\betaspec}{\frac{\diff R}{\diff E}}
\newcommand{\betadoublediff}{\frac{\diff^2 R}{\diff\aem~\diff\eem}}
\newcommand{\betaelossdoublediff}{\frac{\diff^2 N}{\diff\aem~\diff\eem}}
\newcommand{\betaeloss}{\frac{\diff N}{\diff \eem}(\aem)}

\newcommand{\elossfcn}{{f\sub{loss,}}_\aem}
\newcommand{\tofspec}{\frac{\diff N}{\diff t}}
\newcommand{\tofspecbinned}{F(t_j)}
\newcommand{\tof}{\tau}

\newcommand{\smunu}{\sigma\sub{stat}(\mnusq)}

\newcommand{\bkatrin}{b_{0}}

\newcommand{\ftw}[1]{\frac{#1}{12}}

\begin{document}
\title[Neutrino mass sensitivity by TOF spectroscopy]{Neutrino mass sensitivity by MAC-E-Filter based time-of-flight spectroscopy with the example of KATRIN}
\author{Nicholas Steinbrink$^1$, Volker Hannen$^1$, Eric L. Martin$^2$, R. G. Hamish Robertson$^2$, Michael Zacher$^1$ and Christian Weinheimer$^1$}
\address{$^1$ Institut f\"ur Kernphysik, Westf\"alische Wilhelms-Universit\"at M\"unster, Wilhelm Klemm-Str. 9, 48149 M\"unster, Germany}
\address{$^2$ Center for Experimental Nuclear Physics and Astrophysics, and Department of Physics, University of Washington, Seattle, WA, USA}
\ead{n.steinbrink@uni-muenster.de}
\begin{abstract}
The KATRIN experiment aims at a measurement of the neutrino mass with a 90 \% C.L. sensitivity of 0.2 eV/c$^2$ by measuring the endpoint region 
of the tritium $\beta$ decay spectrum from a windowless gaseous molecular tritium source using an integrating spectrometer of the MAC-E-Filter type. 
We discuss the idea of using the MAC-E-Filter in a time-of-flight mode (MAC-E-TOF) in which the neutrino mass is determined by a measurement of the electron time-of-flight (TOF) spectrum that depends on the neutrino mass. 
MAC-E-TOF spectroscopy here is a very sensitive method since the $\beta$-electrons are slowed down to distinguishable velocities by the MAC-E-Filter.
Their velocity depends strongly on their surplus energy above the electric retarding potential.
Using MAC-E-TOF, a statistical sensitivity gain is expected.
Because a small number of retarding-potential settings is sufficient for a complete measurement, in contrast to about 40 different retarding potentials used in the standard MAC-E-Filter mode, there is a  gain in measurement time and hence statistical power.
The improvement of the statistical uncertainty of the squared neutrino mass has been determined by Monte Carlo simulation to be a factor 5 for an ideal case neglecting background and timing uncertainty.
Additionally, two scenarios to determine the time-of-flight of the $\beta$-electrons are discussed, which use the KATRIN detector for creating the stop signal and different methods for obtaining a start signal. 
These comprise the hypothetical case of an `electron tagger' which detects passing electrons with minimal interference and the more realistic case of `gated filtering', where the electron flux is periodically cut off by pulsing the pre-spectrometer potential.
\end{abstract}
\pacs{14.60.Pq, 29.30.Aj}
\submitto{\NJP}
\maketitle


\section{Introduction}

Since the neutrino oscillation results from the Super Kamiokande Experiment in 1998 it is evident that neutrinos have a non-zero rest mass contradictory to the standard model. In a number of experiments on solar, atmospheric, reactor and accelerator neutrinos, the mass differences between the three neutrino mass eigenstates and the three mixing angles have been measured. However, these experiments are not sensitive to the absolute scale of the neutrino mass which needs to be determined in an independent experiment \cite{bib:reactor, bib:solar, bib:atmospheric, bib:longbaseline, bib:oscillations}.

A precise knowledge of the neutrino mass is not only important due to its role in cosmological structure formation~\cite{bib:planck} but also to find out which mass generation mechanism is responsible for the neutrino sector. An important experimental distinction to be made is the determination of the mass hierarchy, i.e. whether the mass states are normally ordered ($m_1 < m_2 < m_3$), inverted ($m_3 < m_1 < m_2$) or quasi-degenerate ($m_1 \approx m_2 \approx m_3$). A determination of the mass hierarchy allows to confirm or rule out models of $\nu$-mass generation. For instance, certain see-saw models, which introduce new physics in the Higgs sector, favour a quasi-degenerate scenario \cite{bib:guintikim}. Moreover, most models introduce additional heavy neutrino mass states, usually identical to or strongly mixed with right-handed (`sterile') flavour states. 
If one of these additional mass states, however, is sufficiently light (i.e. in the eV to keV range) and at least slightly mixed with the 'active' flavour states $\nu_e$, $\nu_\mu$, $\nu_\tau$, the sterile neutrino becomes observable  in neutrino oscillation experiments and in $\beta$ spectra.

Hints towards the existence of light sterile neutrinos in the eV range currently arise for instance from the reactor neutrino anomaly 
\cite{bib:reactoranomaly}, 
the calibration of solar neutrino experiments Gallex and SAGE  \cite{bib:gallex95,bib:gallex10,bib:sage06,bib:sage09} and the short baseline accelerator neutrino oscillation experiments  LSND \cite{bib:lsnd98} and MiniBooNE \cite{bib:miniboone07}. That there might be more than three active neutrinos is also compatible with the total radiation content of the universe obtained from Big Bang nuclei synthesis and the investigation of the Cosmic Microwave Background \cite{bib:steigman12,bib:valentino13,bib:planck}. Sterile neutrinos in the keV range, however, are predicted by the warm dark matter (WDM) scenario, which is supported by several astrophysical observations, e.g. the too less structure at galactic scales (`missing satellite problem')  \cite{bib:wdm}. We will discuss briefly at the end that the method of time-of-flight spectroscopy with gated filter proposed 
in this paper is not only interesting for the search of sub-eV and eV neutrino masses in $\beta$ spectra but also for keV neutrinos.

In probing the absolute $\nu$ mass scale, the Karlsruhe Tritium Neutrino Experiment (\emph{KATRIN}) \cite{bib:designreport} is designed to measure the neutrino mass, which, in the quasi-degenerate regime accessible to KATRIN, is an incoherent sum of the light mass eigenstates weighted with their mixing to the electron neutrino \cite{bib:drexlin2013}
\begin{equation}
\label{eq_effmass}
\mnu = \sqrt{\sum_i \ |U_{ei}^2| \ m_i^2}\ .
\end{equation}

This observable will be determined with a sensitivity of $<$ 0.2 eV/c$^2$ (90 \% C.L.).  The predecessor experiments in Mainz \cite{bib:mainz} and Troitsk \cite{bib:troitsk2011} were successful in finding a 95 \% C.L. upper limit of $\mnu \leq \unit[2.3]{eV}$/c$^2$ and $\mnu \leq \unit[2.05]{eV}$/c$^2$, respectively. A further improvement of the sensitivity beyond the KATRIN design goal would be interesting since the KATRIN sensitivity limit lies in the critical transition region between hierarchical and the quasi-degenerate scenarios and the other methods sensitive to the neutrino mass, the search for neutrino-less double $\beta$-decay \cite{bib:nldb} and cosmology \cite{bib:planck} are of similar sensitivity.

\begin{figure}
\centering
\includegraphics[width=0.6\linewidth]{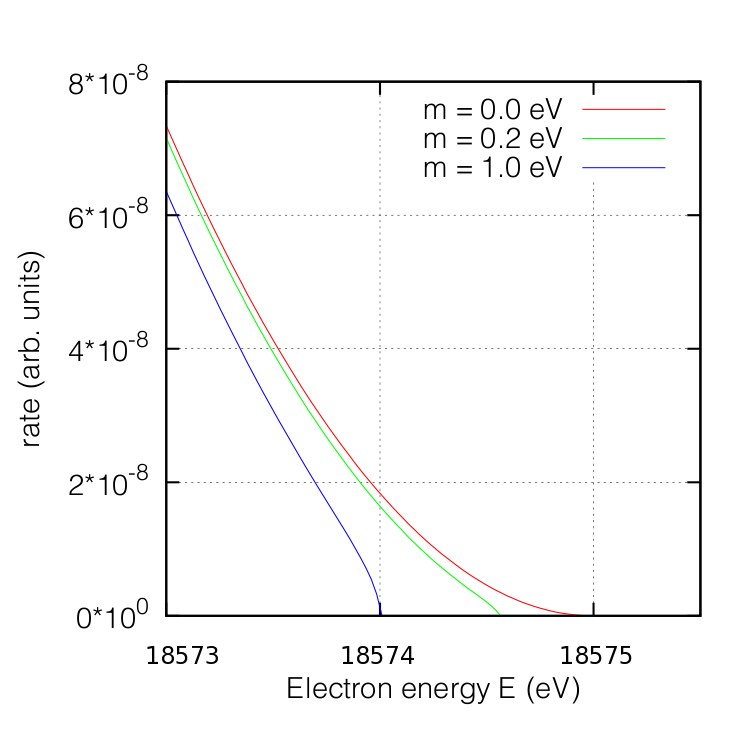}
\caption{Effects of the neutrino mass on the last part of the beta spectrum. The molecular final state spectrum has been omitted for simplicity.}
\label{fig:betaendpoint}
\end{figure}

KATRIN measures the endpoint region of the $\beta^-$ decay spectrum of tritium, which is a function of $\mnu$. Neglecting the recoil energy of the nucleus as well as radiative corrections the differential decay rate \cite{bib:drexlin2013, bib:ottenweinheimer} is at first order given by

\begin{eqnarray}
\label{eq_betaspec}
	\frac{\diff R}{\diff E}(E) & = & N \frac{G_F^2}{2 \pi^3 \hbar ^7 c^5} \cos^2(\theta_C) |M|^2 \ 
	F (E, Z') \cdot p \cdot (E+m_ec^2) \cdot \non \\
	& & \sum\limits_i P_i \cdot (\Eep - V_i - E) \cdot \sqrt{(\Eep - V_i - E)^2 - \mnu^2 c^4} ,
\end{eqnarray}
where $E$ is the kinetic electron energy, $\theta_C$ the Cabbibo angle, $N$ the number of tritium atoms, $G_F$ the Fermi constant, $M$ the nuclear matrix element, $F (E, Z')$ the Fermi function with the charge of the daughter ion $Z^{'}$, $p$ the electron momentum, $P_i$ the probability to decay to an excited electronic and rotational-vibrational state with excitation energy $V_i$ \cite{bib:finalstates,bib:finalstates2, bib:finalstates3} and $\Eep$ the beta endpoint, i.e. the maximum kinetic energy in case of $\mnu = 0$. A non-vanishing neutrino mass changes the spectrum in such a way that the maximal kinetic energy of the decay electrons is lowered to $E\sub{max}=\Eep - \mnu c^2$ and that in the vicinity of the endpoint $E_0 \approx \unit[18.575]{keV}$ the phase space density of the electron is reduced.
Figure \ref{fig:betaendpoint} shows the endpoint region for different neutrino mass values $\mnu$.

KATRIN uses a large spectrometer of MAC-E-Filter type and a windowless gaseous molecular tritium source for the measurement of $\mnu$ \cite{bib:designreport}. The sensitivity of KATRIN 
is principally constrained by the diameter of its spectrometer and its tritium source, which influence the energy resolution and the signal rate, respectively. Since KATRIN reaches the technical limits regarding these parameters, extending the sensitivity would require complementary methods.

In this paper, the idea of a new measurement principle is presented. It can be performed at a suitable MAC-E-Filter setup like the main spectrometer of KATRIN. Instead of the classic integrating mode, where the count rate is scanned as a function of the retarding potential, the time-of-flight (TOF) of every electron passing through the spectrometer is measured. Since the endpoint region of the decay spectrum of tritium is a function of $\mnu$, the distribution of flight times depends as well on the neutrino mass. The MAC-E-Filter TOF mode (MAC-E-TOF) is expected to improve the sensitivity on $\mnu$. Since for each retarding potential not only a count rate but a full TOF spectrum is measured, the number of potential steps can be reduced without sensitivity loss. The measurement time which is gained that way can be invested in obtaining more statistics.

\section{Classic MAC-E-Filter technique}

\begin{figure}
\centering
\includegraphics[width=0.7\linewidth]{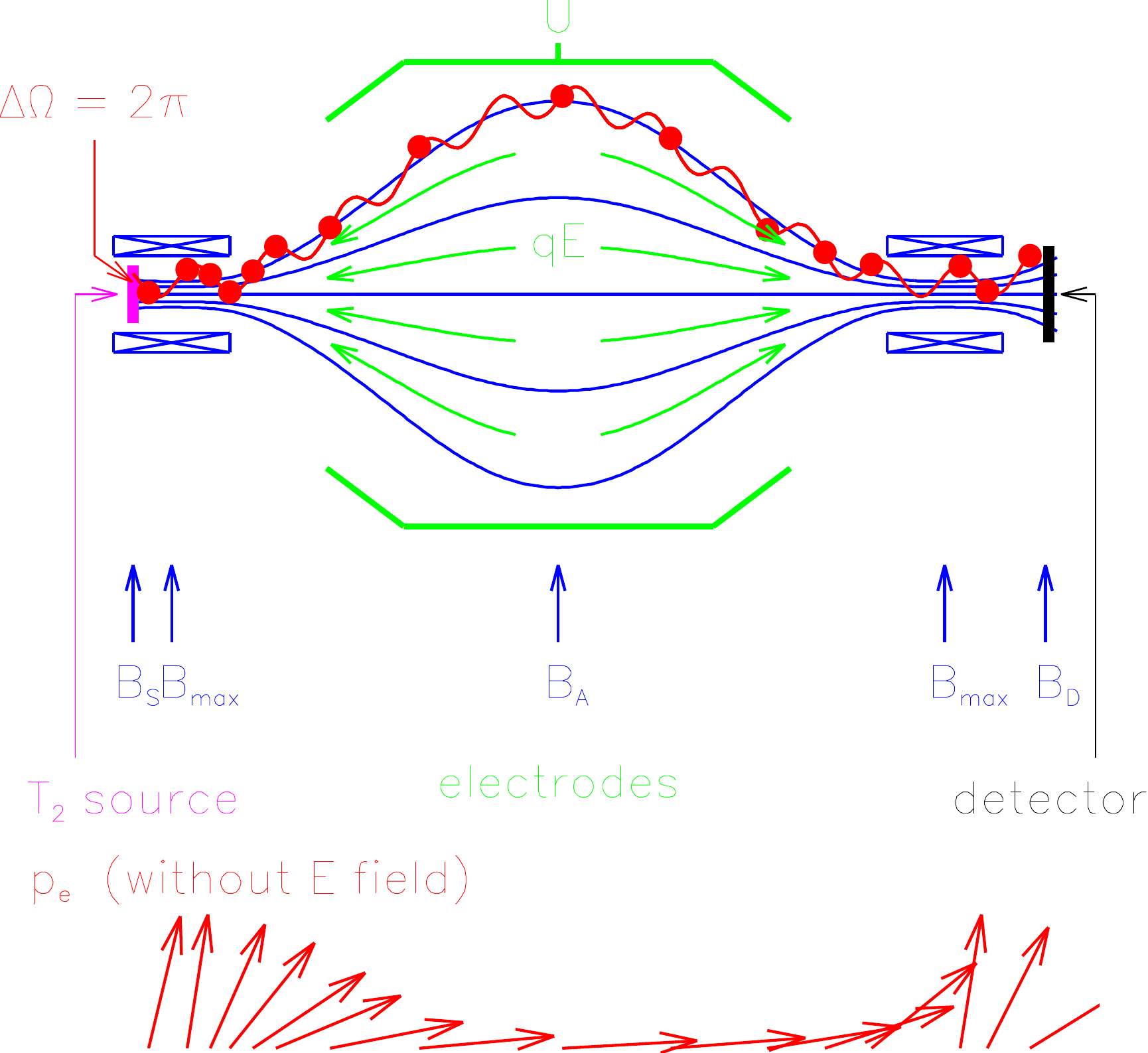}
\caption{Principle of the MAC-E-Filter \cite{bib:designreport}. The transverse momentum is transformed adiabatically into longitudinal momentum. The electron energy is then analyzed by an electrostatic retarding potential.  
}
\label{fig:macefilter}
\end{figure}

 In a classic tritium neutrino experiment the beta spectrum \eqref{eq_betaspec} is scanned with a high pass filter using different threshold energies $qU$. Thereby, the principle of Magnetic Adiabatic Collimation with an Electrostatic Filter (MAC-E-Filter) \cite{bib:picard} is applied. To achieve a sharp energy resolution, the electron  motion which is isotropic at the source is converted into a longitudinal movement in the analysing plane where the retarding potential $qU$ is applied. The transformation is performed by applying high magnetic fields $B_\mathrm{S}$  at the source and $B_\mathrm{D}$ at the detector,  and a low field  $B_\mathrm{min}$ in the analysing plane. Under adiabatic conditions the magnetic moment $\mu$ times the relativistic factor $\gamma$ is conserved,
\begin{equation}
\label{eq_adiab}
	\gamma \mu = \frac{p_\bot^2}{B} = const , 
\end{equation}
meaning that the momentum component transverse to the $B$ field lines $p_\bot$ is converted into parallel momentum in low magnetic field regions (Fig. \ref{fig:macefilter}). 
This parallel beam of electrons is  energetically analyzed by applying the retarding voltage $U$.
  The relative sharpness of this energy high-pass filter depends only on the ratio of the minimum
  magnetic field $B_{\rm min}$ reached at the electrostatic barrier in the so called 
  analysing plane 
  and the maximum magnetic field $B_{\rm max}$ between
  $\beta$-electron source and spectrometer, where $E$ is the starting energy of the electron from an isotropically emitting source:
  \begin{equation} \label{eq:energy_resolution_mace}
    \frac{\Delta E}{E} = \frac{B_{\rm min}}{B_{\rm max}} \ .
  \end{equation}
  
  It is beneficial to place the electron source in a magnetic field $B_{\rm S}$ somewhat lower than    $B_{\rm max}$. 
  Thus the  magnetic-mirror effect  based on the adiabatic invariant (\ref{eq_adiab}) 
  prevents electrons with large starting angles at the source, and therefore long flight paths inside the source, from entering the MAC-E-Filter.
  Only electrons  having starting angles $\theta_{\rm S}$ at  
  $B_{\rm S}$ of:
  \begin{equation} \label{eq:pinch}
    \sin^2(\theta_{\rm S}) \leq \frac{B_{\rm S}}{B_{\rm max}}
  \end{equation}
are able to pass the pinch field $B_{\rm max}$.

The transmission probability $T(E,U)$ of the MAC-E-Filter for an isotropic emitting electron source of energy $E$ can be analytically calculated. Normalized to unity at full transmission it reads:
  \begin{equation} \label{eq:trans}
     T(E,U) = \left\{ \begin{array}{ll} 
          0 & {\rm for~} E \leq qU\\
          \frac{1 - \sqrt{1 - \frac{E-qU}{E} \cdot \frac{B_{\rm S}}{B_{\rm min}}} }{1 - \sqrt{1 - \frac{B_{\rm S}}{B_{\rm max}}} }
              & {\rm for~} qU < E < qU + \Delta E\\
           1  & {\rm for~} E \geq qU + \Delta E 
     \end{array} \right.
  \end{equation}

\begin{figure}
\centering
\includegraphics[width=0.7\linewidth]{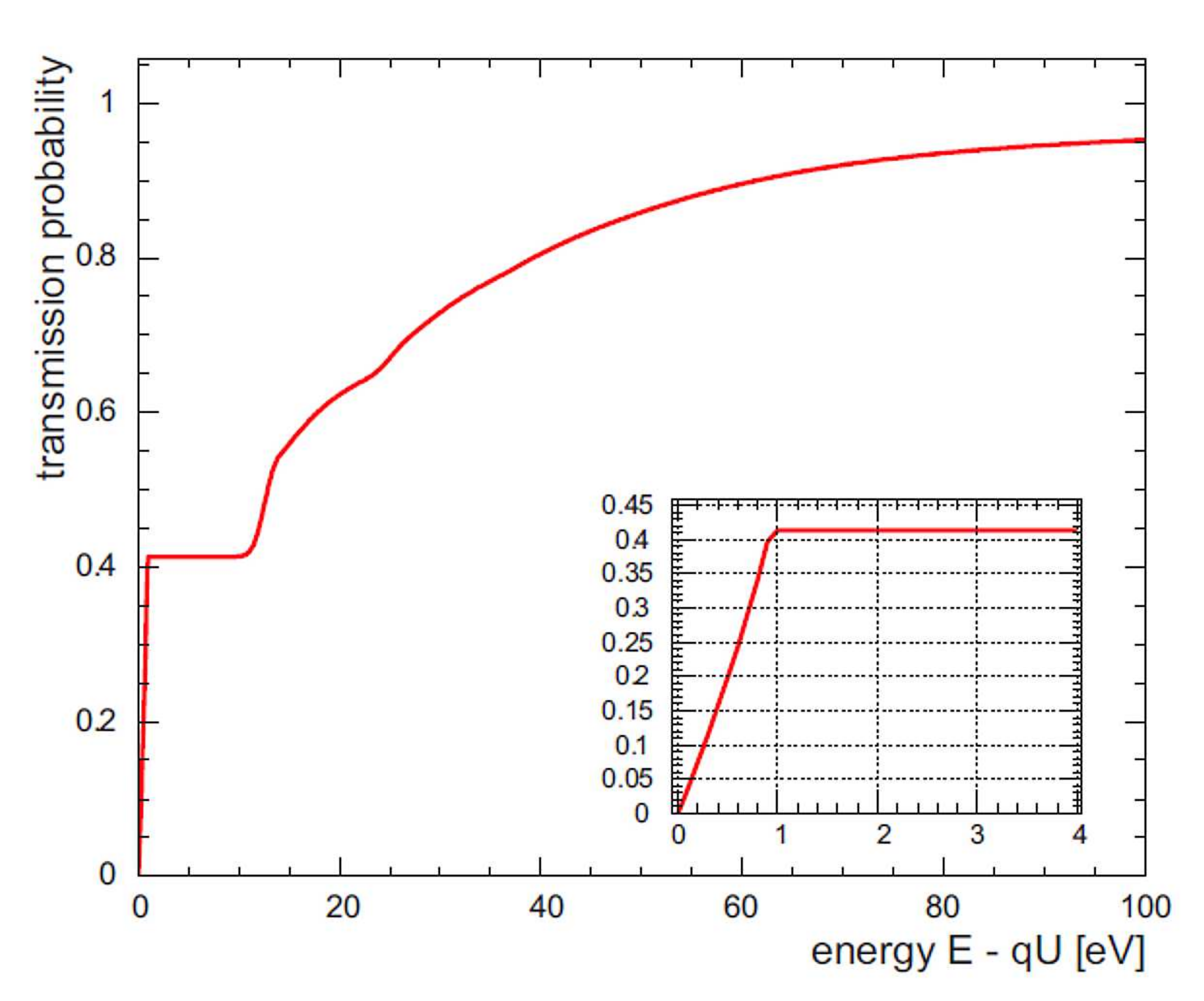}
\caption{Response function of the KATRIN experiment for isotropically emitted electrons \cite{bib:designreport}. The function is given as a convolution of the transmission function $T$ (\ref{eq:trans}) with a function describing the inelastic energy losses in the source \cite{bib:designreport}. The insert shows the rise of the response functions at small surplus energies $E-qU$ governed by the fraction of electrons of approximately 41~\% that have undergone no inelastic scattering process. Therefore this rise corresponds to the shape of the transmission function $T$. 
}
\label{fig:responsefunction}
\end{figure}

In the neutrino mass experiment the count rate is then measured at the detector for each retarding potential $qU$ which is related to \eqref{eq_betaspec} by

\begin{equation}
\label{eq_katrinfitfunc}
R_T(qU) = \frac{\Delta\Omega}{4 \pi}\left( \int\limits_{qU}^{E_0-\mnu c^2} \betaspec(E) \ T^{'} (E,qU) \ \diff E \right) + b \ ,
\end{equation}

where $b$ is the background rate, $\frac{\Delta\Omega}{4 \pi}$ the accepted solid angle with $\frac{\Delta\Omega}{4 \pi} = (1- \cos \theta\sub{max})/2$ and $T^{'}$ the response function of the experiment (Fig. \ref{fig:responsefunction}). The latter is the convolution of the transmission function $T$ (\ref{eq:trans}) with a function describing the inelastic energy losses  \cite{bib:mainz_troitsk_eloss}.

\section{MAC-E TOF spectroscopy}
\subsection{General Idea}

An alternative idea is to use \emph{MAC-E-Filter time of flight (MAC-E-TOF) spectroscopy} to measure the neutrino mass. The time of flight (TOF) of a $\beta$ decay electron through a MAC-E-Filter like the main spectrometer of KATRIN is a function of the kinetic energy and the emission angle. The distribution of the kinetic energies is in first order governed by the beta spectrum \eqref{eq_betaspec} which contains the neutrino mass. By measuring the time of flight distribution (TOF spectrum) of the electrons, one can reconstruct the parameters determining the beta spectrum, including $\mnusq$. Such a method would feature mainly two intrinsic advantages. 

\begin{figure}
\centering
\includegraphics[angle = 270, width = 0.9\linewidth]{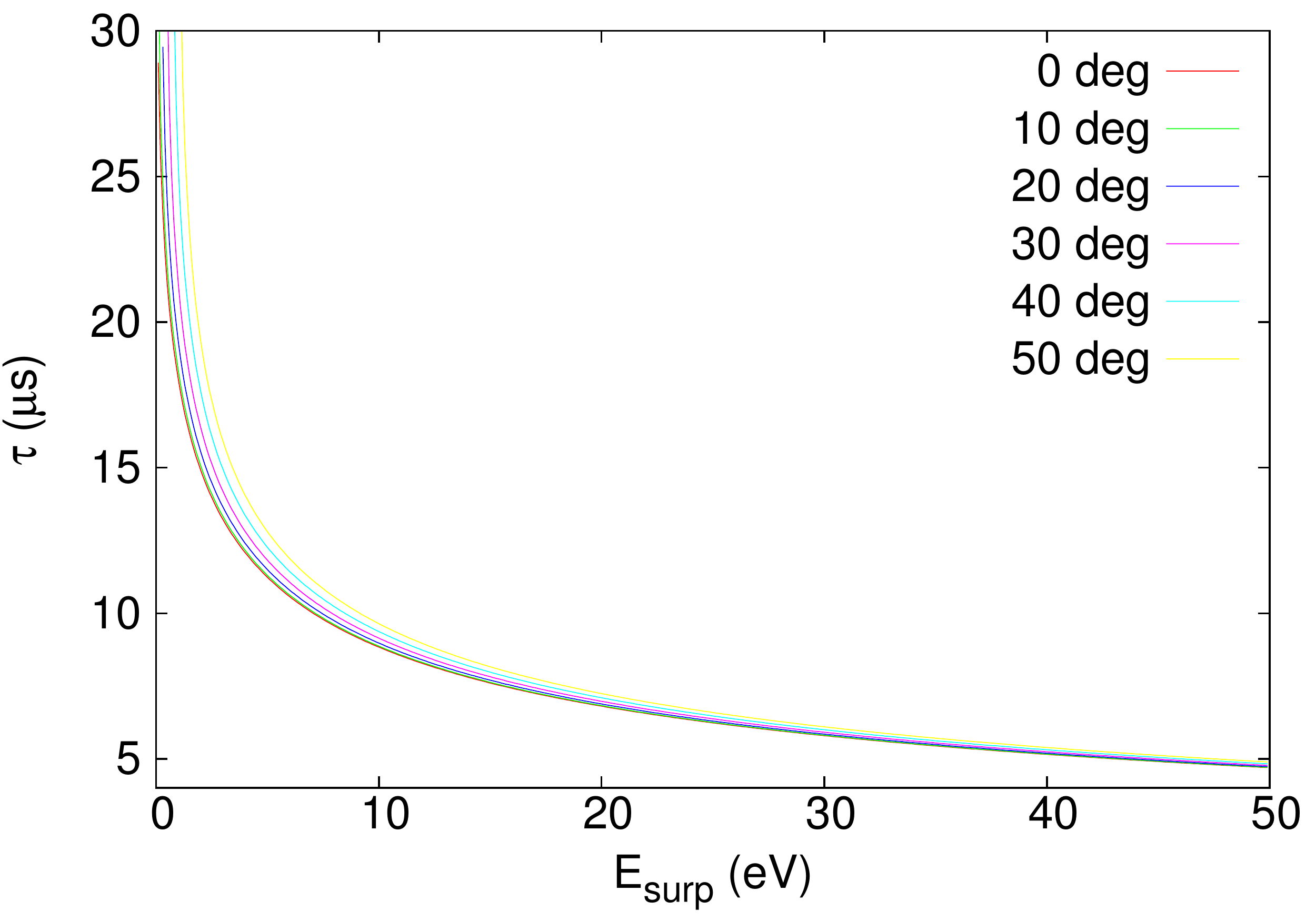}\\
\includegraphics[angle = 270, width = 0.9\linewidth]{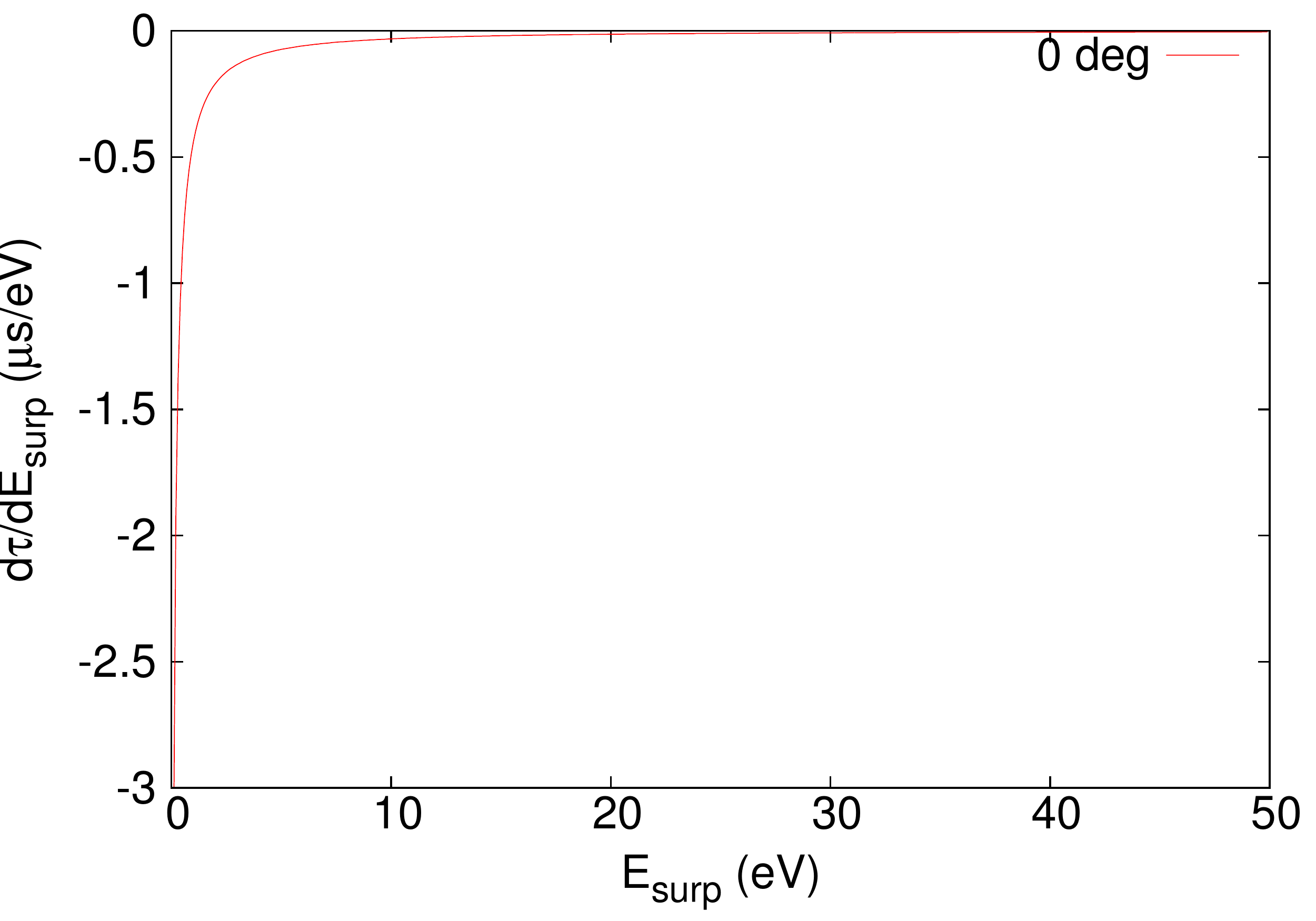}
\caption{Time-of-flight for different starting angles (top) and first derivative for $\theta = 0 ^\circ$ (bottom) as a function of the surplus energy $E\sub{surp} = E - qU$ for a central detector pixel. The starting angle is limited to 50.77 $^\circ$ due to the KATRIN field design. The first derivative reflects the sensitivity on energy differences and is especially large close to the retarding energy $qU$.}
\label{fig:tofsurplusenergies}
\end{figure}

On the one hand the MAC-E-Filter slows down the electrons near the retarding energy. While the relative velocity differences between raw beta decay electrons near the endpoint are tiny, a TOF measurement of beta electrons passing through a MAC-E-Filter will be very sensitive to subtle energy differences just above the retarding energy. It can be seen (Fig. \ref{fig:tofsurplusenergies}) that in principle electron energy differences even below the resolution of the MAC-E-Filter, which is $\Delta E = \unit[0.93]{eV}$ for 18.5 keV electrons in case of KATRIN, can be resolved, given a sufficient time resolution.

On the other hand, the standard MAC-E mode measures only the count rate for each retarding energy, as described above. In contrast, the TOF spectroscopy mode measures the TOF for each decay electron. Thus a full TOF spectrum, sensitive to $\mnusq$, is obtained for each retarding energy. For suitable measurement conditions, this gain of information improves the statistics.

The combination of these advantages allows useful optimizations. In principle it would be sufficient to measure only at a single retarding energy near the beta endpoint, though a small number of selected retarding energies might be more sensitive. Since the systematic uncertainty on $\mnusq$ grows as the retarding energy is decreased, the goal is to minimize the amount of measurements far from the beta endpoint. The TOF method could in principle provide this and concentrate the measurement on a few retarding energies near the endpoint, each of them delivering a full TOF spectrum from which $\mnusq$ can be disentangled.


\subsection{Mathematical Model}

The aim is to state the TOF spectrum as a function of certain fit parameters. These comprise the beta endpoint $E_0$ and the square of the neutrino mass $\mnusq$, as well as the relative signal amplitude $S$ which depends on several factors. Thereby, $E_0$ is in principle known from the $^3$He-T mass difference measurements in penning traps to 1.2 eV precision \cite{bib:endpoint}. However, this is not precise enough to fix it ab initio. Improvements on the beta endpoint precision are on the way \cite{bib:blaum}. For a full study also a constant background rate $b$ needs to be fitted, which is however dependent on the implementation of the measurement method. In order to obtain an expression for the TOF spectrum, the TOF has to be known as a function of the kinetic energy $\eem$ and the starting angle $\aem$ first and then has to be weighted by the corresponding distributions given by the windowless gaseous tritium source \cite{bib:wgts_paper}.

\subsubsection{TOF as function of $\eem$ and $\aem$}

Within the main spectrometer the principle of adiabatic motion, where the magnetic moment is constant \eqref{eq_adiab}, is valid to good approximation. Using a simplified geometry, we take only the field on the $z$ axis into account and neglect magnetron drifts. The $B$ field then is only a function of the $z$ coordinate. Then, the transverse momentum of an electron can be derived from equation 
(\ref{eq_adiab}) as a function of $z$,

\begin{equation}
\label{eq_ptrans}
	p_\bot ^2 (z) = p(\zsource)^2 \cdot \sin^2 \theta (\zsource) \cdot \frac{B(z)}{B(\zsource)},
\end{equation}

where $p(\zsource)$, $\theta (\zsource)$ and $B(\zsource)$ are the electron momentum, its emission angle and the total magnetic field at the electron's starting position $\zsource$. The fraction reflects the role of the adiabatic magnetic field geometry of the MAC-E-filter: As the field $B(z)$ decreases, the transverse momentum is converted continuously into longitudinal momentum. The relativistic energy of the electron is given by the energy-momentum-relation

\begin{equation}
\label{eq_erel1}
	E_{rel}^2(z) = p_\parl^2(z)c^2 + p_\bot^2(z)c^2 + m_e^2c^4.
\end{equation}

Since the total energy $\Etot = \Erel + \Epot = \Ekin + m_e c^2 + \Epot$ is  conserved, 
we can express  the relativistic energy as a function of $z$:
\begin{eqnarray}
\label{eq_erel2}
	\Erel(z) & = & \Erel (\zsource) - \Epot (z) + \Epot (\zsource)   \non \\
	& = & \Ekin (\zsource) + m_e c^2 - q \Delta U(z), 
\end{eqnarray}
where $\Delta U(z)$ is the difference of the retarding voltage at the source and at $z$, 

\begin{equation}
\Delta U(z) =|U(z) - U(\zsource)| \ ,
\end{equation}

and $q$ is the magnitude of the electron charge. Combining Eqs. \eqref{eq_ptrans}, \eqref{eq_erel1} and \eqref{eq_erel2}, we derive an expression for the longitudinal momentum as a function of $z$ only in terms of the field $B$ and the potential difference $\Delta U$:

\begin{eqnarray}
\label{eq_p2c2}
	p_\parl^2(z)c^2 & = & \left(\eem^2 + 2 \eem \ m_e c^2 \right) \left( 1 - \sin^2 \aem \cdot \frac{B(z)}{B(\zsource)} \right) \non \\
	& & + \ q^2 \Delta U^2(z) - 2 q \Delta U(z) \cdot \left( \eem + m_e c^2 \right),
\end{eqnarray}
with the abbreviations $\eem := \Ekin(\zsource)$ and $\aem := \theta(\zsource)$. The TOF is determined by integrating the reciprocal parallel velocity $1/v_\parl = \gamma m/p_\parl = \Erel / p_\parl c^2$ over the measurement path.

\begin{eqnarray}
\label{eq_tof}
	\tof (\eem, \aem) = \int \diff z \ \frac{1}{v_\parl} 
	= \int\limits_{z\sub{start}}^{z\sub{stop}} \diff z \ \frac{\eem + m_e c^2 - q \Delta U(z)}{\sqrt{p_\parl^2(z) c^2} \cdot c}\ .
\end{eqnarray}

The lower bound $z\sub{start}$ of the integration interval depends on where the start signal time is measured{\footnote{This should preferably be at the beginning of the MAC-E-Filter, in case of KATRIN at the entrance of the main spectrometer.}} while $z\sub{stop}$ corresponds to the $z$-position of the detector. As the adiabatic approximation \eqref{eq_adiab} is valid through the whole transport section, this position is arbitrary. The starting angle $\aem = \theta(\zsource)$ is automatically transformed to its correct value at the start position $\theta(z\sub{start})$ by \eqref{eq_p2c2} because only the ratio of local and source magnetic field $B(z)/B(\zsource)$ matters but not the field changes between $B(\zsource)$ and $B(z\sub{start})$.

The integral \eqref{eq_tof} is only correct for electrons emitted in the center of the fluxtube $r=0$ since the integration path is identical with the z axis. As shown in Fig. \ref{fig:macefilter}, the electrons perform a cyclotron motion around the B field lines, where for the flight-times only the velocity component parallel to the B field lines ${v_\parl}$ needs to be considered. Electrons emitted at $r=0$ take a path different from the z axis. This is, however, not a real shortcoming of our method because at KATRIN the starting position can be reconstructed by the point of arrival on the multipixel detector. Therefore we do not
expect signicant changes in sensitivity depending on the electron emission radius. For this  in-principle study of the statistical sensitivity we consider it to be sufficient to use the central electron tracks only.

\subsection{TOF Spectrum}

Equation \eqref{eq_tof} presumes a fixed kinetic starting energy and starting angle as arguments. For a real source, these parameters are not fixed but follow physical distributions. In order to be calculated numerically, the differential TOF spectrum $\tofspec$ is discretized into bins of constant length $\Delta t$ and integrated over each bin $j$, leading to a binned spectrum

\begin{equation}
\label{eq_def_tofbinned}
	\tofspecbinned := \int\limits_{t_j}^{t_{j+1} = t_j + \Delta t} \diff t \ \tofspec \ .
\end{equation} 

The number of events in a certain TOF bin depends on the distribution of starting energies and angles $\eem$ and $\aem$,

\begin{eqnarray}
\label{eq_tofspec}
	\tofspecbinned 
	& = & 
	{ {\int\int} \atop { \mbox{\small $(\eem , \aem)$ with $t_j \leq \tof(\eem , \aem) \leq t_{j+1}$}}}
	\quad \betaelossdoublediff ~\mathrm d\aem~\mathrm d\eem  \non\\
	& = & 
	\int\limits _0 ^{\amax}	\
	\int\limits_{\eem_j(\aem)}^{\eem_{j+1}(\aem)}
	 \quad \betaelossdoublediff ~\mathrm d\aem~\mathrm d\eem
\end{eqnarray} 


where $\betaelossdoublediff$ is the double differential event rate as function of $\eem$ and $\aem$. The integral limits $E_j(\aem)$ and $E_{j+1}(\aem)$ are defined in such way that $\tof(E_j, \aem) = t_j$ and $\tof(E_{j+1}, \aem) = t_{j+1}$, respectively. At first order, $\betaelossdoublediff$ is given by the double differential decay rate $\betadoublediff$ into the accepted solid angle $\Delta \Omega \over 4 \pi$,

\begin{equation}
\label{eq_firstorder}	
	\betaelossdoublediff \approx \betadoublediff \ .
\end{equation}

As the double differential is proportional to the joint probability distribution of emitting an electron with energy $\eem$ at a polar angle of $\aem$ and, furthermore, the angle and the energy are uncorrelated in case of a non-oriented radioactive source, the quantity can be separated into a product of the single differential decay rate $\betaspec$, as given by \eqref{eq_betaspec}, and the angular probability distribution $\angledist$ \cite{bib:cowanstatistical},

\begin{equation}
\label{eq_doublediff}	
	\betadoublediff = \betaspec \cdot \angledist\ .
\end{equation}

In case of an isotropic tritium source, a sine law applies for the angular distribution,

\begin{equation}
\label{eq_angledist}	
	\angledist = {1 \over 2} \sin \aem \ .
\end{equation}


This angular distribution function is normalized to unity over the full solid angle $4\pi$. Since for KATRIN the polar angle is restricted to $\amax = 50.77^\circ$, the signal rate is implicitly reduced by a factor

\begin{equation}
\label{eq_anglenorm}	
	\int\limits _0 ^{\amax} d\theta \ \angledist = \frac{\Delta \Omega}{4\pi} = \frac{(1 - \cos\amax)} 2
\end{equation}

which is enforced by the upper integral bound $\amax$ in \eqref{eq_tofspec}.

The approximation \eqref{eq_firstorder} is only valid in case of an ideal tritium source. However, quite a few electrons lose energy in elastic and inelastic scattering processes with the tritium molecules. These losses are dependent on the emission angle since the path through the tritium source increases with $1/cos \theta$. Thus, for the differential rate of events which are actually analysed in the main spectrometer, given by $\betaelossdoublediff$ in \eqref{eq_tofspec}, starting energies and angles become correlated. Additionally, the signal rate decreases due to several losses inside the experiment. A factor $\epsilon\sub{flux} \approx 0.83 $ applies since the flux tube transported through the whole system corresponds to a diameter of $82$ mm w.r.t to the beam tube diameter of 90~mm, meaning that only a part of the WGTS tube is imaged onto the detector. Furthermore, the detector efficiency gives an additional factor of $\epsilon\sub{det}\approx 0.9$.

In total, the true event rate can be calculated by by applying the correction factors and convoluting the beta spectrum with an energy loss function, which gives

\begin{eqnarray}
\label{eq_betaeloss}
	\betaelossdoublediff 
	& = & 
	\epsilon\sub{flux}\cdot\epsilon\sub{det} \cdot \angledist \cdot \betaspec \otimes \elossfcn \non\\
	& = & 
	\epsilon\sub{flux}\cdot\epsilon\sub{det} \cdot \angledist \cdot \left(
	p_0(\aem) \cdot  \betaspec 
	+ 
	\sum\limits_{n=1}^\infty p_n(\aem) \cdot \betaspec \otimes f_n \right), \non\\
\end{eqnarray}
where the $f_n$ is the energy loss function of scattering order $n$ which is defined recursively through the single scattering energy loss function $f_1$ as

\begin{equation}
\label{eq_eloss}
f_n =  f_{n-1} \otimes f_1 \qquad (n > 1) \ .
\end{equation}

The function $f_1(\Delta E)$ is the probability density of losing the energy $\Delta E$ in a singular scattering event  \cite{bib:mainz_troitsk_eloss}. The functions of $f_n$ can then correspondingly be interpreted as the same for $n$-fold scattering. In this equation, all changes of the angle of the electron during scattering are neglected. $p_n$ is the probability that an electron is scattered $n$ times. If we again neglect changes of the angle, it is a function of the emission angle $\theta$ and given by a Poisson law

\begin{equation}
\label{eq_pscat}
	p_n(\theta) = \frac{\lambda^n(\theta)}{n!}e^{-\lambda^n(\theta)}~.
\end{equation}
Here, the expectation value $\lambda$ is given in terms of the column density $\rho d$, the mean free column density $\rho d\sub{free}$ and the scattering cross section $\sigma\sub{scat}$ as
\begin{equation}
	\lambda(\theta) = \int_0^1 \mathrm d x \ \frac{\rho d \cdot x}{\rho d\sub{free} \cdot \cos \theta} = \int_0^1 \mathrm d x \ \frac{\rho d \cdot x \cdot \sigma\sub{scat}}{\cos \theta}~,
\end{equation}
where the integration factor $x$ accounts for the fact that the starting position of the electron inside the WGTS is statistically distributed.

\section{Simulation of principle}

\subsection{Study of TOF spectra}

\begin{figure}
\centering
\includegraphics[angle = 270, width = 0.49\linewidth]{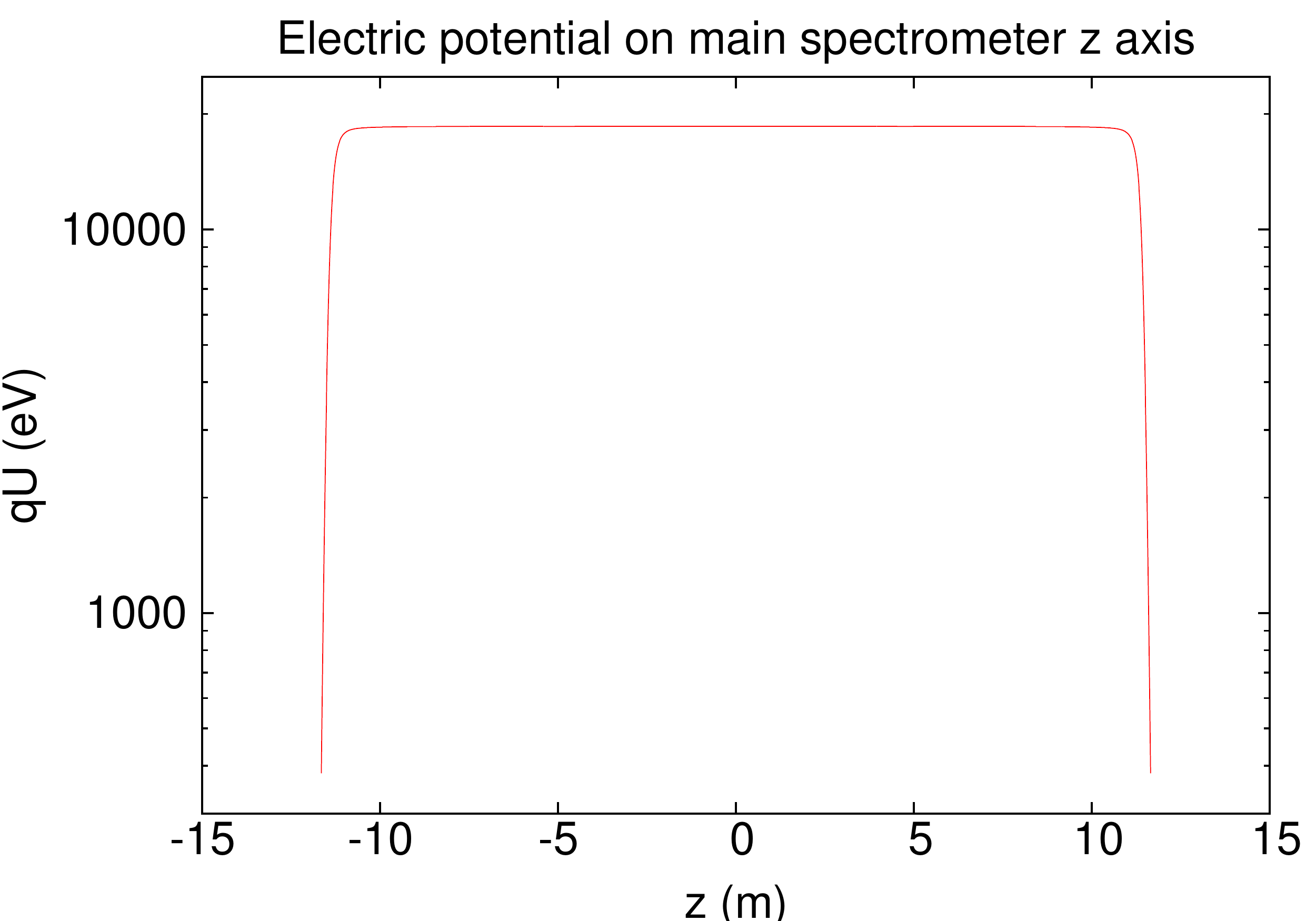}
\includegraphics[angle = 270, width = 0.49\linewidth]{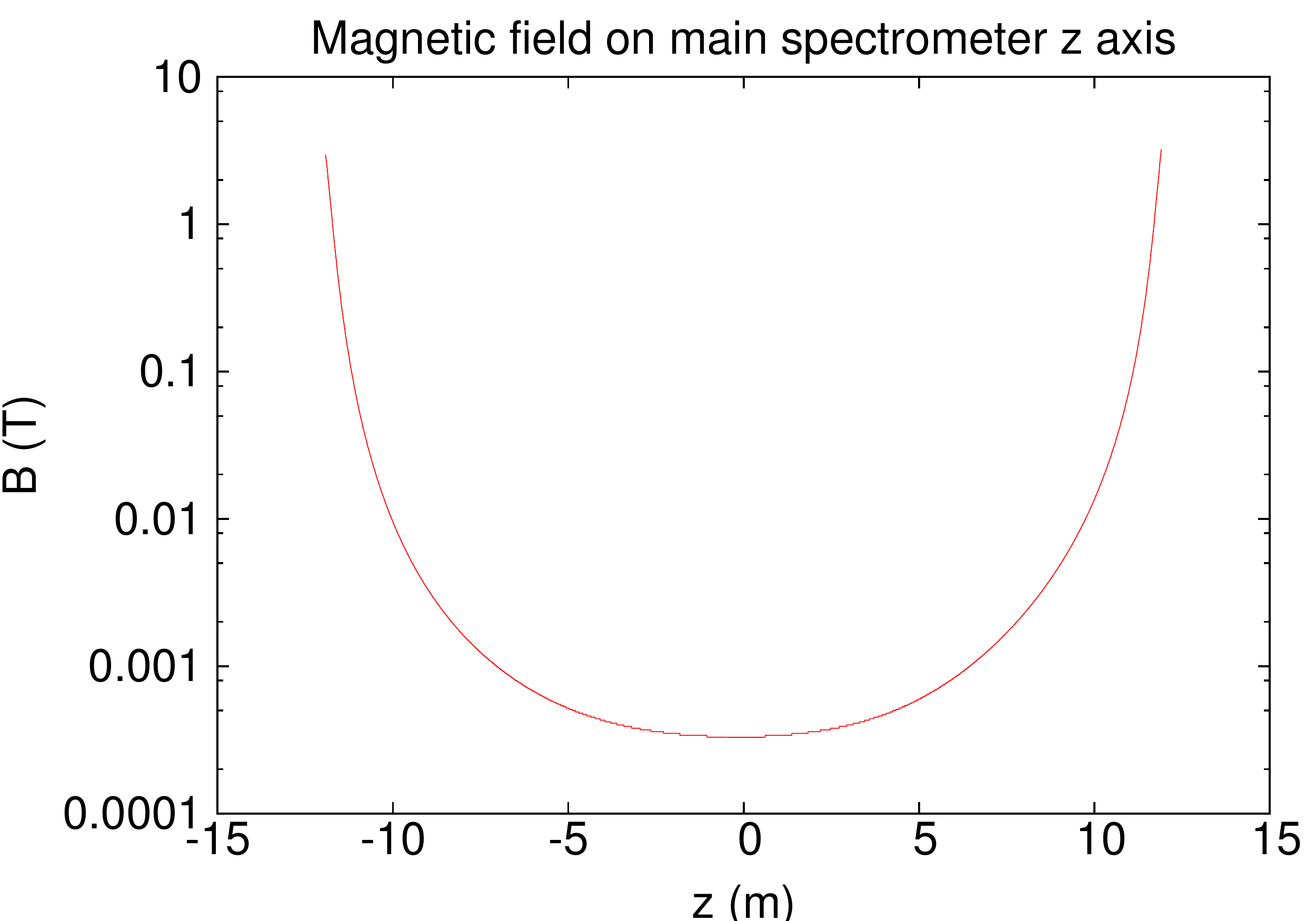}
\caption{Electric potential (a) and magnetic field (b) along the inner axis of the KATRIN main spectrometer generated by simulation. The scaling of the electric potential depends on the retarding energy $qU$.}
\label{fig:fieldmaps}
\end{figure}

To calculate the TOF spectrum according to \eqref{eq_tofspec} the input parameters have to be obtained. A model for the one-dimensional field maps $\Delta U(z)$ and $B(z)$ in \eqref{eq_tof} has been determined by the KATRIN simulation tools \emph{magfield} and \emph{elcd3\_2} \cite{bib:kathrinphd} using a modestly simplified geometry that contains the most important coils and electrodes in the main spectrometer (Fig. \ref{fig:fieldmaps}). A model for the energy loss function \eqref{eq_eloss} has been determined in the past 
by electron scattering experiments on hydrogen \cite{bib:mainz_troitsk_eloss} and refined by using
excitation and ionisation data from hydrogen molecules \cite{bib:glueck_eloss}. The final-state excitation spectrum of the daughter molecules has been used from reference \cite{bib:finalstates}.

\begin{figure}
\centering
\includegraphics[width=0.9\linewidth, clip = true, trim= 0 0 0 22.5]{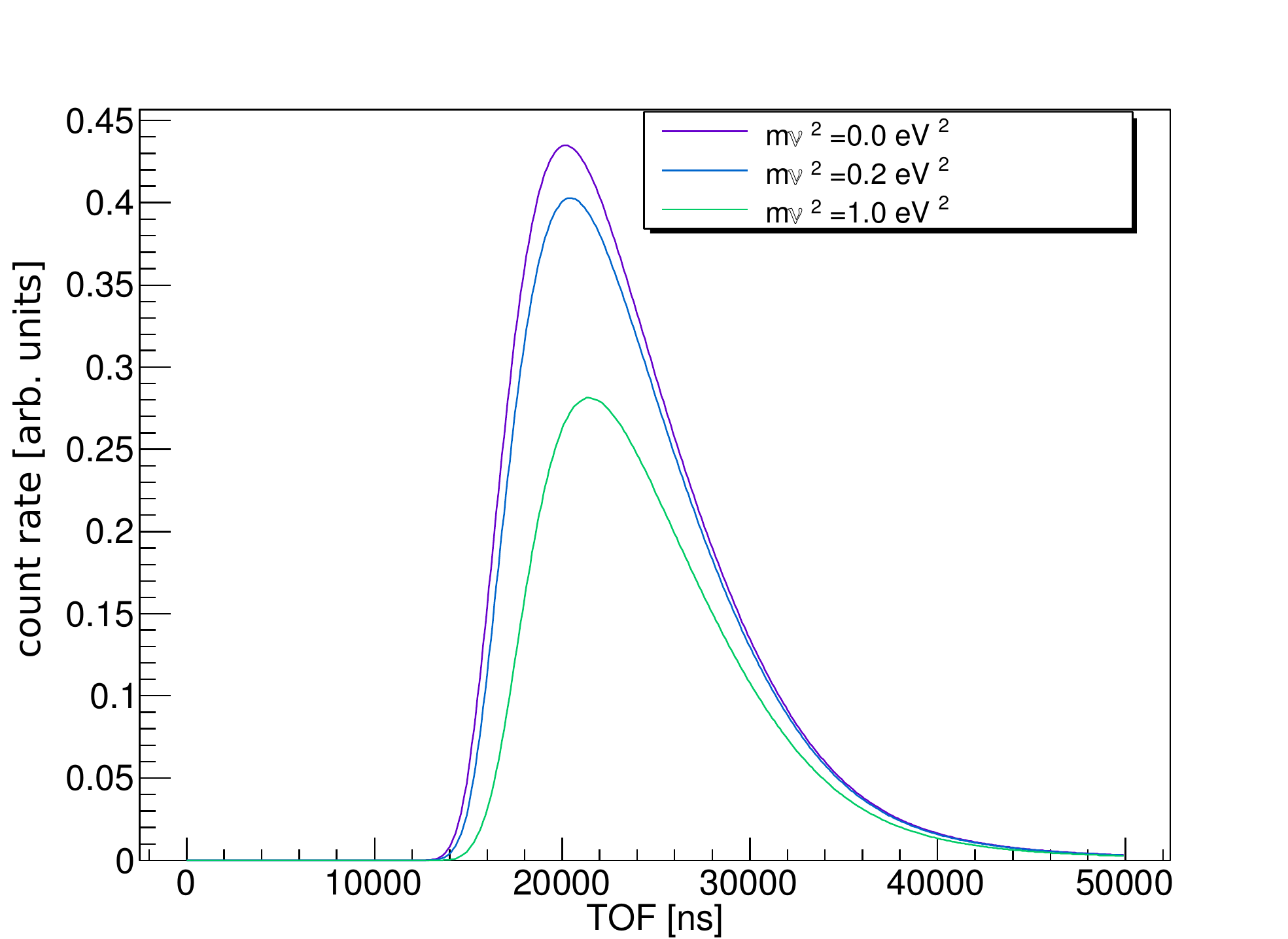}
\caption{Effects on the TOF spectrum for different neutrino masses at a high retarding potential (18570 eV) with endpoint $\Eep =\unit[18574.0]{eV}$. The scaling of the y-axis is arbitrary.}
\label{fig:numasseshighpotential}
\end{figure}

A typical set of simulated TOF spectra for different neutrino mass squares is shown in Fig. \ref{fig:numasseshighpotential}. The following details and parameter-dependent behaviour can be observed:

\begin{itemize}
\item For each spectrum, there exists a minimal TOF $t\sub{min}$. This corresponds to the maximum kinetic emission energy that an electron can have, given by $E\sub{max} = \Eep - \mnu c^2$. 
\item From $t\sub{min}$ on, a steep slope begins, leading soon to a maximum somewhat above $t\sub{min}$, followed again by a long, slow fall. The maximum can be explained by the fact that the higher the energy becomes, the lower the number of electrons is, due to the shape of the end of the beta spectrum, whereas the 'TOF energy density', i.e. the interval size of the energy that corresponds to a certain TOF bin, increases. These effects balance each other, leading to a maximum somewhere in the middle.
\item There is no maximal TOF. The closer the energy of an electron is to the retarding energy, the slower it will be. That means that electrons with an energy infinitesimally above the retarding energy will have an infinite TOF.
\item If the neutrino mass square $\mnusq$ is changed, the main signature is a change in count-rate and a change of the shape especially at the short-time end of the spectrum (Fig. \ref{fig:numasseshighpotential}). 
\item A higher retarding energy $qU$ leads to a clearer distinction between spectra for different neutrino masses. The reason is that the neutrino mass is mainly visible in the last few eV of the beta spectrum. Therefore, it seems  optimal for the TOF mode to measure with retarding energies near the endpoint. However, due to the lower count rate and the difficult decorrelation of neutrino mass square and endpoint, measurements from lower retarding energies should be added to the data.
\end{itemize}

\subsection{Neutrino mass fits}

\subsubsection{Method} 

In order to study the statistical uncertainty we used the spectra to fit Monte Carlo (MC) data. The MC data have been obtained by creating Poisson distributed random numbers based on the predictions from \eqref{eq_tofspec}, where certain choices of the parameters $\mnusq$ and $\Eep$ as well as the retarding energy $qU$ and the measurement time have been assumed. The data are fitted in this self-consistent method by the models \eqref{eq_tofspec} using a $\chi^2$ minimization method. If multiple measurements with different $qU$ are assumed, they can be fitted with a common chi square function by adding the chi square functions from each run. If the fit is performed correctly, the chosen parameters $\mnusq$ and $\Eep$ are reproduced. Additionally, estimates for the parameter errors can be determined as
\begin{equation}
\label{eq:errorbars}
\chi^2 (\phi_0 \pm \Delta \phi_\pm) = \chi^2 (\phi_0) + 1 \ ,
\end{equation}
where $\phi_0$ is a parameter estimate and $\Delta \phi_\pm$ are the requested, not necessarily symmetric parameter error bars \cite{bib:cowanstatistical}. To obtain a symmetric $\chi^2$ parabola for neutrino masses near zero, there must also be an extension for a negative $\mnusq$ that joins smoothly with the physical spectrum for $\mnusq > 0$. To accomplish this, to each term in the sum of the beta spectrum \eqref{eq_betaspec} a factor
\begin{equation}
 	f_i = \left( 1 + \frac{\meff}{\epsilon_i} e^{-(1+\epsilon_i/\meff)} \right)
\end{equation} 
is applied in case of $\mnusq < 0$ and $\epsilon_i + \meff > 0$. In this expression, the abbreviations $\epsilon_i = E_0 - V_i - E$ and $\meff = \sqrt{-\mnusq}$ have been used \cite{bib:chefmainzpaper}. This method allows a simple but realistic prediction of the statistical uncertainty of $\mnusq$.

\subsubsection{Results}

In order to determine the improvement potential by the TOF mode, an optimal choice of the measurement times of the runs with different retarding energies $qU$ has to be made. For KATRIN a total on-line time of three years is planned, which has to be distributed among the retarding energies. We used a simple algorithm where we discretized the retarding potentials and the measurement time and determined the statistical uncertainty with the method above for all possible permutations. The results are shown in table \ref{tb_measopt}. At the MC data creation, a neutrino mass of zero has been assumed. In this case, the average of the fit uncertainties $\mnusq$, as given by \eqref{eq:errorbars}, describes the sensitivity on the neutrino mass squared.

\begin{table}
\small
\caption{\label{tb_measopt}Average statistical uncertainty $<\smunu> ~=~ <\frac{1}{2} (|{\Delta\mnusq}_-| + | {\Delta\mnusq}_+|)>$ (arithmetic mean of positive and negative error of 10 simulations and fits), average fit parameters $<\bar\Eep>$ and  $<\bar\mnusq>$, as well as Pearson's correlation coefficient $R(\Eep,\mnusq)$ of uniform and optimized distributions and the KATRIN standard mode. The total assumed measurement time is the KATRIN standard of three years \cite{bib:designreport}, distributed among four retarding energies $qU=$18550 V, 18555 V, 18560 V and 18565 V as well as for single retarding potentials for $\mnusq = \unit[0]{eV^2/c^2}$, $\Eep = \unit[18575]{eV}$ and $b=0$. The choice of retarding potentials for the TOF mode is motivated by the idea that a choice of a few potentials close to the endpoint will likely improve the systematics additionally to the statistical uncertainty.}
{\footnotesize 
\begin{tabular}{@{}lllllll}
\br
fraction of measure- 		& distribution 	& lowest		& mean						& mean fitted		& mean fitted						& mean\\
ment time per 				& type 				& retarding 	& stat. error				& endpoint			& $\nu$ mass squared			& correlation\\
retarding energy 			&  						& energy 		& $<\smunu>$ 			& $<\bar\Eep>$		& $<\bar\mnusq>$				& coefficient \\
 									&						& {[eV] } 		& {[eV$^2$/c$^4$] } 	& {[eV] }	& {[eV$^2$/c$^4$] }	&R($\Eep$,$\mnusq$) \\
\mr
$(\ftw 3, \ftw 3, \ftw 3, \ftw 3)$  	& uniform   & 18550 & 0.0033 & 18574.9997 & 0.0004 & 0.65\\
$(\ftw 1, \ftw 0, \ftw 3, \ftw 8)$  	& optimized & 18550 & 0.0032 & 18575.0002 & 0.0013 & 0.70\\
$(0, \ftw 4, \ftw 4, \ftw 4)$  			& uniform   & 18555 & 0.0034 & 18575.0002 & 0.0015 & 0.73\\
$(0, \ftw 2, \ftw 1, \ftw 9)$  			& optimized & 18555 & 0.0034 & 18575.0002 & 0.0006 & 0.72\\
$(0, 0, \ftw 6, \ftw 6)$  					& uniform   & 18560 & 0.0036 & 18575.0002 & 0.0014 & 0.74\\
$(0, 0, \ftw 4, \ftw 8)$  					& optimized & 18560 & 0.0035 & 18575.0007 & 0.0034 & 0.76\\
\mr
$(1, 0, 0, 0)$  & single & 18550 & 0.0035 & 18575.0000 & 0.0004 & 0.82\\
$(0, 1, 0, 0)$  & single & 18555 & 0.0036 & 18575.0000 & 0.0003 & 0.88\\
$(0, 0, 1, 0)$  & single & 18560 & 0.0038 & 18574.9999 & -0.0015 & 0.79\\
$(0, 0, 0, 1)$  & single & 18565 & 0.0039 & 18574.9998 & 0.0007 & 0.66\\
\br
- & standard mode & 18555 & 0.020 & \\
- & standard mode & 18550 & 0.019 & \\
- & standard mode & 18545 & 0.018 & \\
\mr
\end{tabular}}
\end{table}

The comparison shows that it is in principle sufficient to measure at only one retarding energy. 
If this single retarding energy is close to the endpoint, the correlation between the parameters $\Eep$ and $\mnusq$
becomes weaker at the cost of losing count-rate. It turns out that it is beneficial to combine measurements at
more than one retarding energy, where this relation between lowest retarding energy and correlation coefficient does not neccessarily hold true (see table \ref{tb_measopt})  In almost all tested cases using multiple retarding potentials a solid decorrelation without suffering from too little count-rate has been possible. 

The results in Table \ref{tb_measopt} correspond to the optimum case since background, time uncertainty and other limitations have been neglected. They reflect the maximal improvement potential that can be achieved with a TOF mode. The motivation to neglect the background in the optimum case is based on the idea that a sensitive TOF measurement method may be able to reduce the background, too, depending on the implementation of the measurement.  It can be shown that, compared with the statistical sensitivity for the reference configuration of KATRIN, $\smunu = \unit[0.018]{eV^2/c^4}$, an improvement of up to a factor $\sim$ 5-6 
by TOF spectroscopy is possible. The actual improvement factor, however, depends on the method by which the time-of-flight determination is implemented.

Furthermore, one can conclude that even a reduction of the systematic uncertainty  might be possible with the TOF mode. This is due to the fact that the systematic uncertainty at KATRIN depends heavily on the measurement interval at which the spectrum is scanned \cite{bib:designreport}. That is mainly caused by the uncertainty of the parameters of the electron energy loss which becomes more important at lower retarding energies. An ideal TOF mode, in contrast, would allow one to measure solely at higher retarding energies. 

\begin{figure}
\centering
\includegraphics[width=\linewidth]{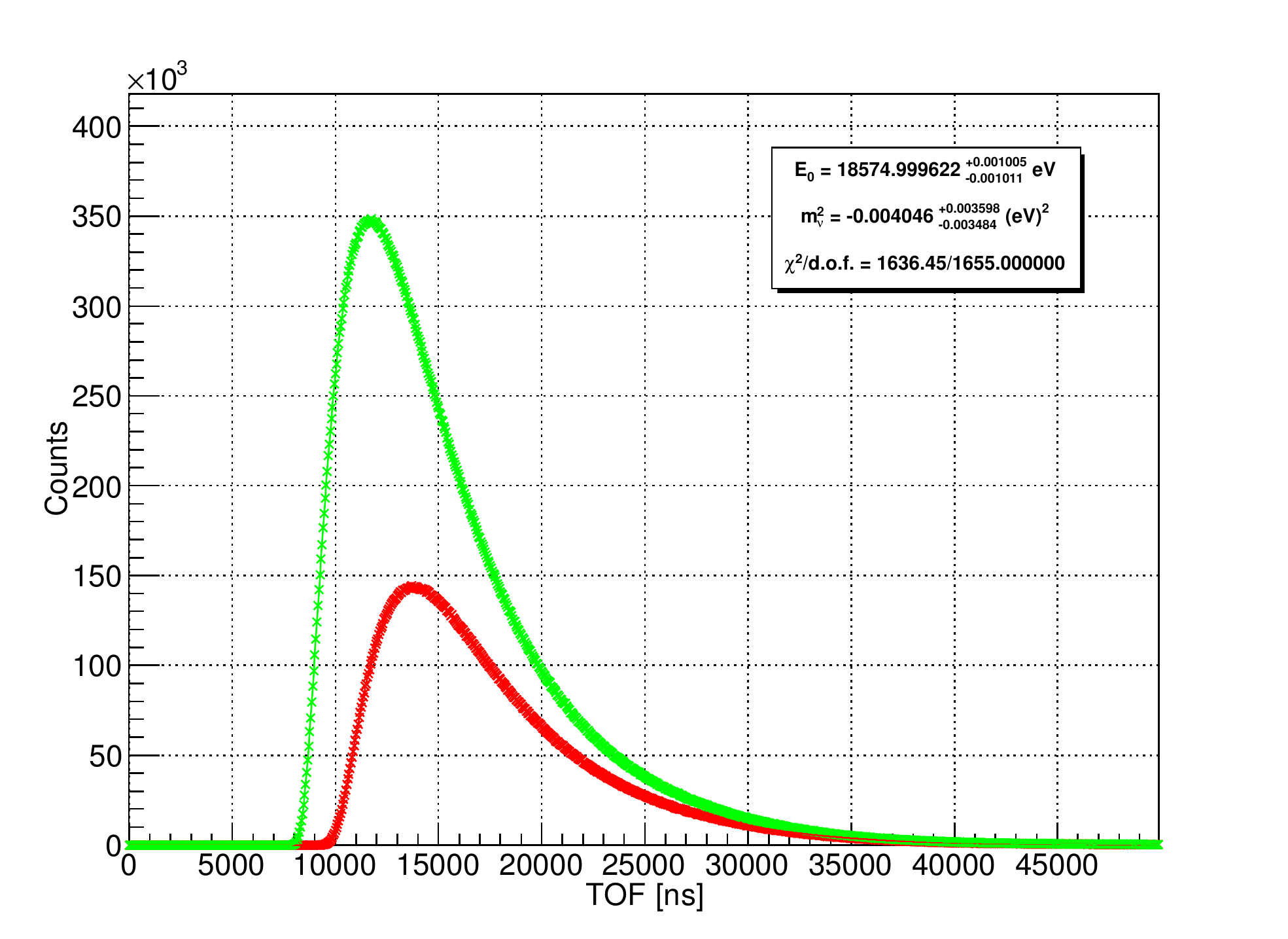}
\caption{Example of simulated data and fit of a TOF spectrum for the optimal case of no background and no time uncertainty. For the fit, a measurement time of 3 years was assigned $2/3$ to 18565 eV (red points online, smaller amplitude) and $1/3$ to 18560 eV (green points online, larger amplitude).   These correspond to the optimum distribution for  lowest retarding energy of 18560 eV in Table \ref{tb_measopt}).}
\label{fig:optimumexample}
\end{figure}

For further analyses, the optimal distribution for the case of 18560 eV lowest retarding energy has been taken which is likely a good compromise between statistical and systematic uncertainty. An example for a fit, based on this measurement time distribution, is shown in Fig. \ref{fig:optimumexample}.

\section{Principles of electron tagging}

The creation of a TOF spectrum requires a start and a stop time associated with the passage of each detected electron through the spectrometer.   The stop time is easily derived from the pulse produced when the electron is absorbed in the focal-plane detector.   A start time must be produced by some device upstream from the spectrometer.  For clarity in the following discussion the start and stop times are defined in this way, although in practice an electronic clock or digitizer record would probably be started with the ``stop'' time signal in order to increase the  live time. 

In addition to the time interval $\tau$ between start and stop, two other time intervals enter the problem.  One is the time interval $\delta\tau$ for the electron to pass through the start detector, and the second is the time resolution $\Delta\tau$.  We will assume for brevity that both of these are entirely associated with the start-time detection process.

The objective is to detect the electron without perturbing its motion to such an extent that its original energy can no longer be reliably measured in the spectrometer.  An amount of energy $\Delta E$ must be extracted from the electron as it flies through the start detector in order to derive a trigger signal.  The uncertainty in the start time is related to the uncertainty introduced in the electron's energy by
\begin{equation}
\Delta E \Delta \tau \ge  \hbar.
\end{equation}
Since $\hbar = 6.58 \times 10^{-7}$ eV~ns, it is clear that the uncertainty principle is no impediment to a TOF measurement.  This remains true even if one replaces $\Delta \tau$ by the much smaller quantity $\delta \tau$, arguing conservatively that the energy uncertainty should be related to the available observation time in the start detector, a time of order a few ns, instead of the desired time-of-flight uncertainty, a time of order 100 ns. 

Another, somewhat more stringent, criterion is that sufficient energy must be extracted from the electron to produce a signal that cannot be mistaken for a random fluctuation of thermal noise:
\begin{equation} \label{eq:kboltzmann}
\Delta E  \gg  k_\mathrm{B}T.
\end{equation}
Since $k_\mathrm{B}=8.62 \times 10^{-5}$ eV K$^{-1}$, this condition presents no in-principle impediment to a TOF measurement either. On the other hand we do not want to change the energy
of the tagged electron too much. Therefore equation (\ref{eq:kboltzmann}) calls for a low-temperature detection device.

Notwithstanding these favorable indications, it is difficult to extract sufficient energy from a single  electron moving in an apparatus of any size larger than microscopic.  There are four methods that can be applied:
\begin{enumerate}
\item Radiation from an electron undergoing accelerated motion,
\item Work done by image charges moving through a load circuit, 
\item Work done by magnetically induced currents flowing through a load circuit, and
\item Interaction of electrons with other electrons, e.g. in atoms.
\end{enumerate}  
We consider each of these approaches in turn.

\subsection{Radiation from an electron undergoing accelerated motion}

For low-energy or mildly relativistic electrons moving in a magnetic field, the dominant energy-loss mechanism is cyclotron radiation~\cite{bib:monreal}.  The power radiated scales as the square of the magnetic field $B$, and depends on the pitch angle $\theta$ with respect to the static field.  The cyclotron angular frequency is
\begin{eqnarray}
\omega&=& \frac{qB}{\gamma m_e} \equiv \frac{\omega_0}{\gamma}
\end{eqnarray}
and the radiated power is
\begin{eqnarray}
P(\beta,\theta) &=& \frac{1}{4\pi \epsilon_0} \frac{2q^2 \omega_0^2}{3c} \frac{\beta^2 \sin^2{\theta}}{1-\beta^2},
\end{eqnarray}
where $q$ is the charge on the electron, and $m_e$ is the electron mass.   At a field strength of 4.5 T, as is found between the KATRIN pre-spectrometer and the main spectrometer, the radiated power for 18.6 keV electrons having a 51 degree pitch angle is about 8 fW.  If the observation region has a usable length $x_m$, the time spent by the electron in that region is
\begin{eqnarray}
\delta \tau = \frac{x_m}{c\beta \cos \theta}
\end{eqnarray}
which, for $x_m = 50$ cm, ranges from  6 to 10 ns depending on the pitch angle.   The energy lost by the electron in free radiation during its transit of the observation region is then $\le 10^{-3}$ eV, only $12~k_\mathrm{B} T$ at 1K. The cyclotron frequency $\omega/2\pi$ is 126 GHz, still a technically  challenging region to work in with non-bolometric amplifiers, and the antenna collection efficiency will necessarily be a compromise in order to allow the electrons to make an unobstructed transit. 

More specifically, the noise power is $k_\mathrm{B}T\Delta\nu$, where $\Delta\nu$ is the bandwidth.  There are two principal contributions to the bandwidth -- the line broadening caused by the duration $\delta \tau$ of the signal, and the broadening caused by magnetic-field inhomogeneity.   We assume the latter can be made smaller than the former, and neglect it.  Only a fraction $A$ of the radiated signal energy is absorbed in the receiver.  The signal power is distributed over a bandwidth $\Delta\nu=(2\pi\  \delta\tau)^{-1}$.  Hence the energy from the signal $W_s$ in the interval $\delta\tau$ may be compared with the noise energy $W_n$ as follows:
\begin{eqnarray}
W_s &=& A P(\beta,\theta)\   \delta\tau \\
W_n &=& k_\mathrm{B}T\Delta\nu \  \delta\tau \\
\frac{W_s}{W_n} &=& \frac{2\pi AP(\beta,\theta) \  \delta\tau}{k_\mathrm{B}T}.
\end{eqnarray}
Taking $\delta\tau = 10$ ns, the bandwidth is 16 MHz and the signal-to-noise ratio in that interval is about 16 when $\theta=51$ degrees, $T=1$K, and $A = 0.5$.  The magnetic-field inhomogeneity must be less than about $10^{-4}$.   The signal-to-noise ratio drops rapidly below 51 degrees pitch angle, which calls for a much longer magnet of high homogeneity.

Up to now we discussed classical electromagnetic radiation but we have to consider the quantisation as well. The energy of a cyclotron radiation photon $E_\mathrm{ph} = \hbar \omega$ amounts to
$5 \cdot  10^{-4}$ eV, which is comparable to the whole emitted radiation energy. This again calls for more radiated energy $\Delta E$ and thus for a  longer magnetic field section.

\subsection{Work done by image charges moving through a load circuit}

Schottky pickups encompass a broad class of devices used to detect particle beams without intercepting them.  They have in common the induction of an image charge in a circuit connected to plates or a cavity.   To estimate the energy that can be extracted from a single electron, we consider a conducting cylinder that the electron enters, passes through, and emerges from.   The image charge induced on the cylinder gives it a stored energy
\begin{eqnarray}
W &=& \frac{q^2}{8\pi\epsilon_0 r}
\end{eqnarray}
where $r$ is the radius of the probe ring.  For $r=5$ cm, the stored energy is about $10^{-8}$ eV, which can be delivered to a suitably matched external circuit.   Because the observation region $x_m$ is considerably longer than the radius of the aperture, several probes can be cascaded to increase the energy extracted.   If the probes are cascaded in a phased way, some further enhancement of the energy extraction can be achieved, but the dependence of electron transit times on pitch angle limits the usefulness of resonant structures.

\subsection{Work done by magnetically induced currents flowing through a load circuit}

The electron in flight produces a time-varying magnetic field that can be used to induce a transient current in a nearby inductively-coupled element.  The magnetic field can be obtained from a Lorentz transformation on a stationary electron, 
\begin{eqnarray}
B(\zeta,t) & = & \frac{\mu_0}{4\pi}\frac{\gamma\beta c q \zeta}{(\zeta^2+\gamma^2\beta^2c^2t^2)^{3/2}} 
\end{eqnarray}
where $\zeta$ is the impact parameter, or distance of closest approach, and $t$ the time since closest approach.   For the present purposes, we neglect the cyclotron motion because its period is two orders of magnitude shorter than the characteristic time for establishing the magnetic  field due to average linear motion.  The magnetic energy in the field beyond a minimum radius $\zeta=r_i$ is 
\begin{eqnarray}
W &=& \pi \int_{r_i}^\infty \int_{-\infty}^\infty B^2  \beta \zeta c\  d\zeta\  dt. \\
&=&  \frac{3\pi^2}{4}  \left(\frac{\mu_0}{4\pi}\right)^2 \frac{\gamma \beta^2 c^2 q^2}{r_i}
\end{eqnarray}
The integral over time is equivalent to integrating over the third, axial, coordinate.   Assuming all of the magnetic energy outside $r_i$ can be delivered to an external circuit, the energy extracted from the electron is $1.6 \times 10^{-15}$ eV for $r_i = 5$ cm.   The estimate is somewhat pessimistic, because the spatial width of the instantaneous magnetic field distribution is considerably narrower than $x_m$, and so this amount of energy could be extracted several times with multiple inductors, but the available energy is so small that the details are not important. 

\subsection{Interaction of electrons with other electrons}

This approach is standard and useful, but the energy extracted is not under control for any given collision, and sometimes may be much larger than desirable.   Recently it has been proposed~\cite{bib:coldatoms} to use highly excited Rydberg atoms as an interaction medium, since the cross section for modest energy losses is enhanced relative to more inelastic collisions. 

\subsection{Electron-tagger random triggers}

In addition to the statistical advantage in measuring the signal, a substantial gain in background suppression can also be expected with electron tagging.  The background suppression is based on the principle that a signal in the detector is placed in delayed coincidence with a signal from the tagger, within a time window  of width $t_o$.   Given a sufficiently low expected rate of tagger signals $r_s$, and neglecting pile-up, random coincidences would result in  background reduction by a factor $\phi$ with 

\begin{equation}
\frac{1}{\phi} =  1 - e^{-t_o \cdot r_s} \ ,
\end{equation}

Too high a rate of tagger signals, either due to a high flux of incoming electrons or to a high noise level, would impair the measurement. However, in a dual-spectrometer setup like KATRIN, the electron flux through the tagger can be reduced by the pre-spectrometer down to $\mathcal{O}$($10^{3}$ Hz). On the other hand, electrons can be trapped between the pre- and main spectrometer, and would give rise to a high rate of tagger signals. Trapped electrons, however, could be reduced by an active measure such as a scanning wire \cite{bib:beck} or suitably chosen time-dependent perturbations of the electric fields along the transport path.

The tagger must be placed before the main spectrometer, but should
be placed in a low rate area. For best tagger functionality, it is also favorable to have the beam
tube through the tagger fill a small area. The highest-field region  of
the 4.5 T magnet between the pre-spectrometer and main spectrometer is
a good position; there magnetic reflection reduces the flux through the tagger even below the rate at the pre-spectrometer analyzing plane. There are two major sources of electrons at this
point in the beam line.
\begin{itemize}
\item Electrons from the source that make it through the pre-spectrometer,
are reflected by the retarding potential at the main spectrometer
analysing plane, and then go back through the pre-spectrometer and are
absorbed in the source section. These will each pass the electron tagger
twice.
\item Electrons from the source that make it through the pre-spectrometer,
are reflected by the main spectrometer, but, because of radiative losses or scattering inside the main spectrometer, lack the energy to make
it back through the pre-spectrometer and become trapped.  The rate of electrons becoming trapped is only $\mathcal{O}$($1$ Hz),  but, without any active removal method each trapped electron could pass the electron tagger millions of times before becoming undetectable.  An active removal method only needs to remove trapped electrons in $\mathcal{O}$($10^{4}$ passes) before the rate from trapped electrons is lower than the rate from electrons reflected by the main spectrometer retarding potential.
\end{itemize}

\subsection{Random rate from pre-spectrometer}

Electrons from the source that pass both the pre-spectrometer retarding
potential, and the magnet, each make two passes.

From equation \eqref{eq:pinch} applied to the pre-spectrometer, we find
the magnet at the end of the pre-spectrometer reflects all electrons
with a starting angle at the source greater than:

\begin{equation}
\theta_{\rm pre}=\arcsin(\sqrt{\frac{B_{\rm source}}{B_{\rm max}}}),\label{eq:thpre}
\end{equation}
where $B_{\rm max}$ is the magnetic field in the magnet
at the end of the pre-spectrometer.

Solving equation \eqref{eq_p2c2} for $E$ and requiring that the parallel
momentum $p_\parallel$ must be positive to make it past
the retarding potential, we find the minimum initial kinetic energy
for passage through the pre-spectrometer is given by:

\begin{equation}
\label{eq:emin}
E_{\rm min}(\theta,\triangle U,B)=
\frac{\sqrt{k q^2 \Delta U^2 + (k - 1)^2 m_e^2c^4 } +q \Delta U + (k - 1)mc^2}
{1-k}
\end{equation}

\[
k=\sin^{2}\theta\frac{B(z\sub{pre})}{B(z\sub{source})},
\]
where $\theta$ is the starting angle at the source, $\Delta U$
is the retarding potential of the pre-spectrometer, and $B(z\sub{pre})$ is the magnetic
field at the analysing plane of the pre-spectrometer. 

\begin{figure}
\centering
\includegraphics[width=0.75\linewidth]{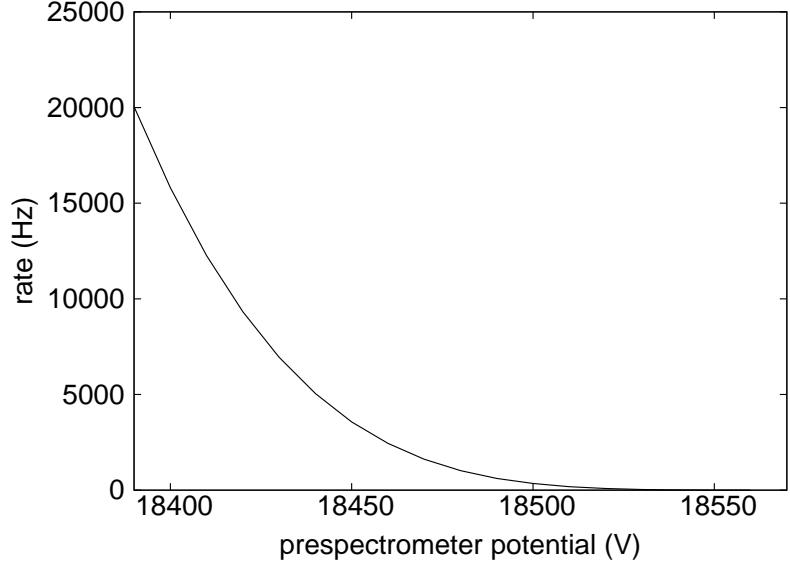}

\caption{Rate of electrons passing the electron tagger, due to electrons which
make it through the pre-spectrometer, are reflected in the main spectrometer,
and return through the pre-spectrometer to be absorbed in the
source section.}
\label{fig:noiseratefromprespec}
\end{figure}

\begin{figure}
\centering
\includegraphics[width = 0.75\linewidth]{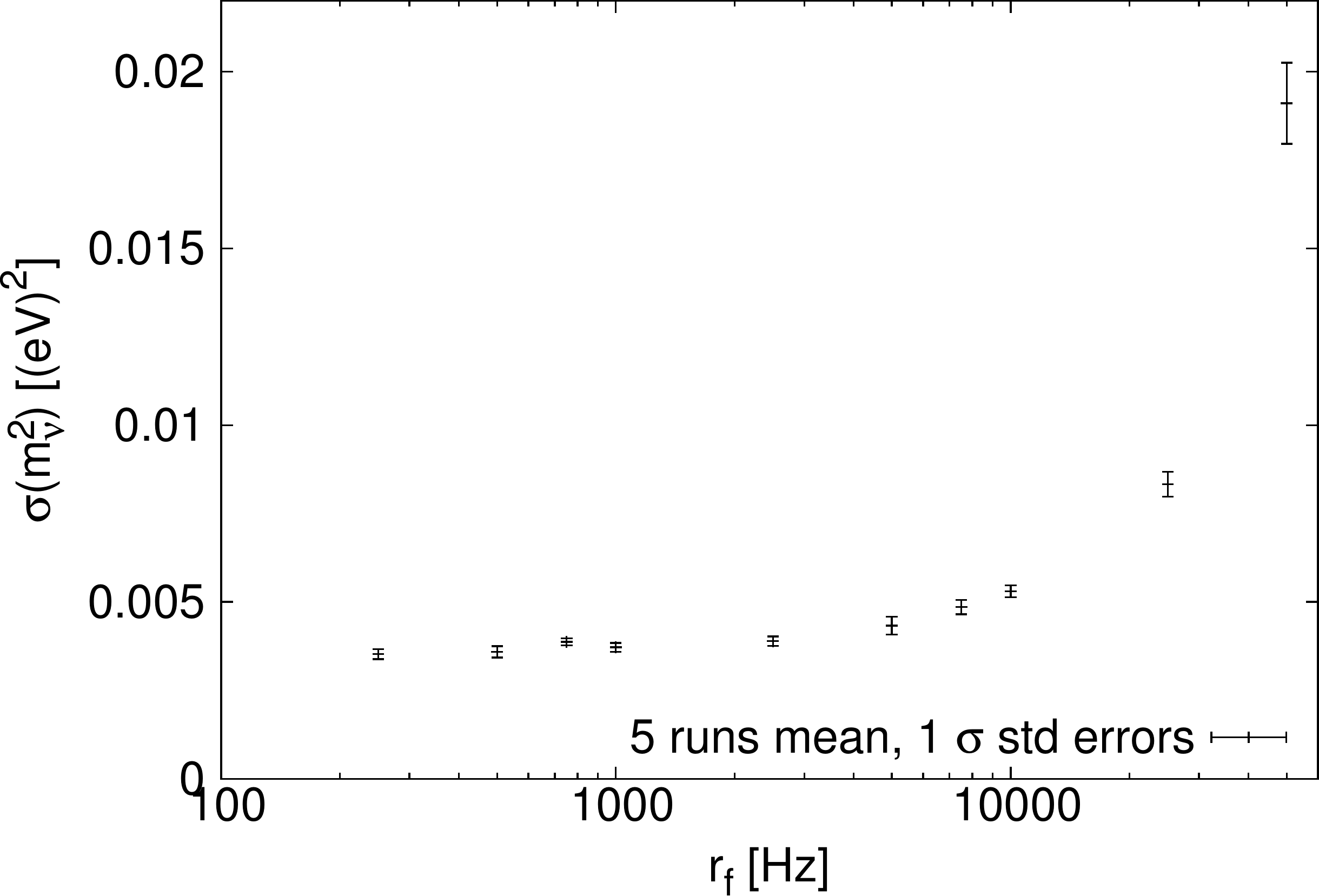}
\caption{Statistical uncertainty of $\mnusq$ as a function of tagging rate $r_f$ which is given by $R\sub{pre}$ in eq. \eqref{eq:prespecrate}, in case no other source of tagging events is present. The results are based on a measurement time distribution of $2/3$ and $1/3$ of three years in total, assigned to 18565 eV and 18560 eV, respectively (optimum distribution for 18560 eV lowest retarding energy in Table \ref{tb_measopt}). For each point the results from five simulation runs with identical parameters and different random numbers have been averaged. Other sources of background and time resolution have been neglected.}
\label{fig:taggingrates}
\end{figure}

Combining equations \eqref{eq:thpre} and \eqref{eq:emin} we find the
average rate of electrons from the pre-spectrometer passing the electron
tagger is:

\begin{equation}
\label{eq:prespecrate}
R\sub{pre} = 2\intop_{0}^{\theta_{\rm pre}}\frac{2\pi\, \sin\:\theta}{4\pi}d\theta\intop_{E_{\rm min}}^{E_{0}}\betaeloss,
\label{eq:prespnoiserate}
\end{equation}
where $\betaeloss$ is the rate at the source, given by equation
\eqref{eq_betaeloss}. The factor 2 has been applied since each electron makes two passes.
This assumes that  all electrons that make it through the pre-spectrometer
are reflected by the main spectrometer, of which the vast majority are.  The rate as a function of pre-spectrometer potential is plotted in Fig.~\ref{fig:noiseratefromprespec}. In Fig.~\ref{fig:taggingrates} the neutrino mass sensitivity as a function of the rate at the tagger is plotted. It can be seen that rates below $\sim$ 10 kHz do not cause a significant loss of sensitivity.

\subsection{Prospects for single-electron tagging}

In summary, there is no fundamental obstacle to the detection of single electrons in flight, if the specification is only that energy perturbations be small enough to be acceptable for reliable energy determination.  However, as a practical matter, no satisfactory method has yet been  identified that allows a controlled and sufficiently large amount of energy to be extracted from the electron.  The cyclotron-emission method may be adequately sensitive, but significantly longer superconducting magnets providing a very homogeneous field are needed at the tagger position, which are not available between the KATRIN spectrometers.  With a successful tagger, there is the potential for suppression of backgrounds by virtue of the coincidence requirement between the electron tagger and the focal-plane detector,  but  there is also a random trigger rate for the tagger caused by electrons that do not pass through the main spectrometer.

\section{Simulation of the measurement method}

We turn now to numerical simulations in order to evaluate the performance of a TOF system under generic assumptions.

\subsection{Generic parameters}

\begin{figure}
\centering
\includegraphics[width = 0.75\linewidth]{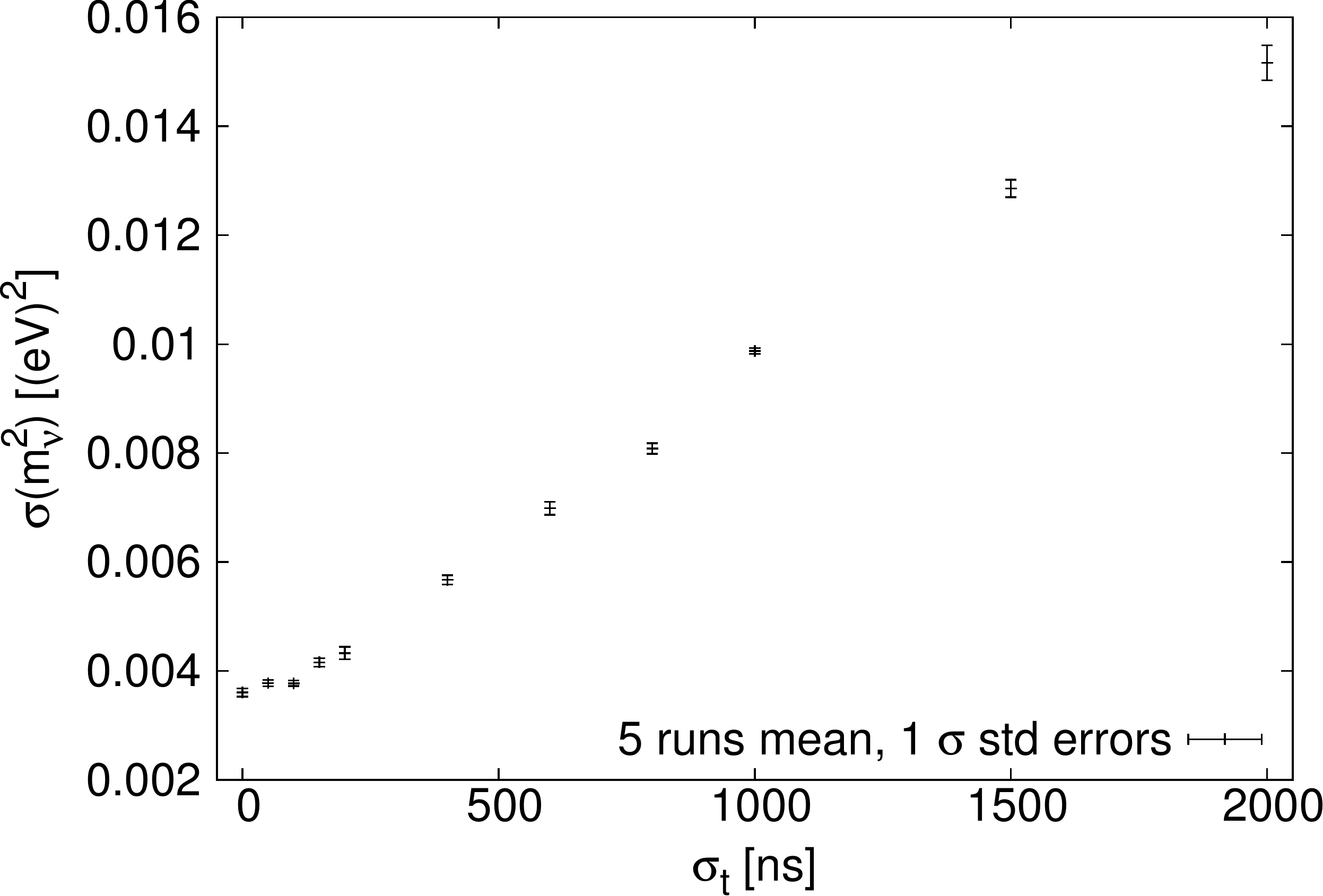}
\includegraphics[width = 0.75\linewidth]{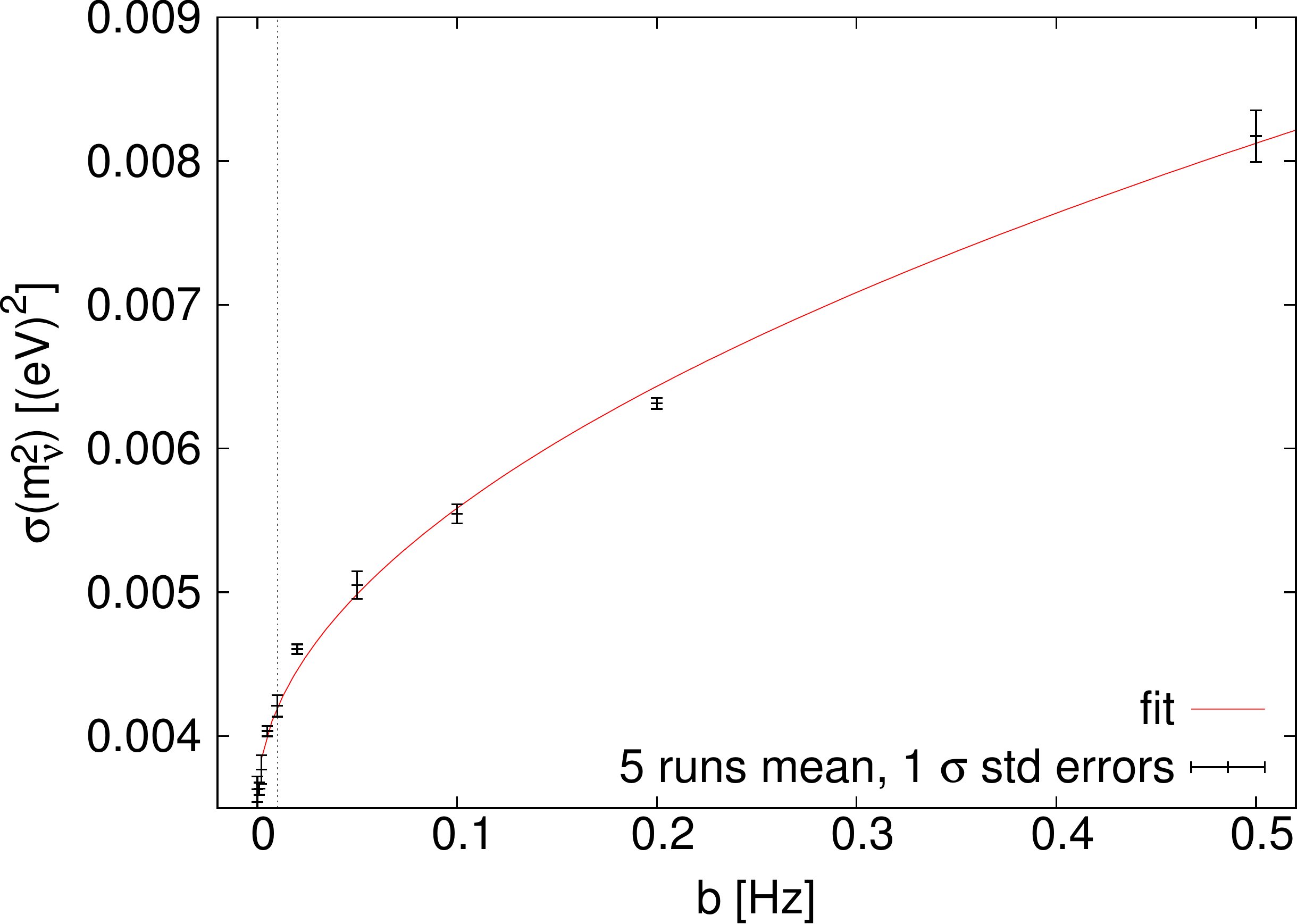}
\caption{Statistical uncertainty of $\mnusq$ as a function of time resolution $\sigma_t$ (a) and background rate $b$ (b). The results are based on a measurement time distribution of $2/3$ and $1/3$ of three years in total, assigned to 18565 eV and 18560 eV, respectively (optimum distribution for 18560 eV lowest retarding energy in Table \ref{tb_measopt}). In both plots the results from 5 simulation runs with identical parameters and different random numbers have been averaged. The dashed line in (b) corresponds to the KATRIN standard mode design goal of $b$ = 10 mHz. The solid line in (b) represents the best fit by an inverse power law of the form $\sigma(x) = \sigma_0 + x^{1/a}$.}
\label{fig:genericparameters}
\end{figure}

Several parameters apply to most methods, chiefly the background rate, the time resolution and the efficiency. The dependence of the statistical uncertainty on the efficiency $\epsilon$, i.e. the ratio of events whose TOF is correctly measured, follows a $1/\sqrt{\epsilon}$ law. This behavior is theoretically predicted and has been verified by simulations. The dependence on the background rate and the time resolution found in the simulations is shown in Fig. \ref{fig:genericparameters}.

For the time resolution a Gaussian uncertainty has been assumed. Fig. \ref{fig:genericparameters} shows that the timing is uncritical for resolutions within the order of magnitude of the KATRIN detector. For resolutions in the range up to 200~ns the error increases by about 20 \%. The scale of this behaviour is plausible as the scale of the neutrino-mass-sensitive part of the TOF spectrum is mainly contained in the first few $\mu$s after the onset (see Fig. \ref{fig:numasseshighpotential}) and becomes washed out if the time resolution of the TOF measurement method exceeds some 100 ns.

Assuming the background level in the TOF mode is the same as in the standard mode where it is specified for KATRIN as $b < \unit[10]{mHz}$, it can be shown that the improvement by the TOF mode is still up to a factor 3 in terms of $\mnusq$. The behaviour follows a power law with approximately  $\smunu =  0.006~\mathrm{eV^2/c^4} \cdot (b/\mathrm{Hz})^{1/2.0} + 0.004~\mathrm{eV^2/c^4}$.\footnote{Here, no correlation between background and starting signal as in the tagger case has been assumed.} In comparison, in the standard mode the background dependence can be determined to be approximately  $\smunu =  0.019~\mathrm{eV^2/c^4} \cdot (b/\mathrm{mHz})^{1/1.7} + 0.009~\mathrm{eV^2/c^4}$ \cite{bib:susanne},   in reasonable agreement with the analytically approximated formula of $\smunu \propto b^{1/3}$ \cite{bib:ottenweinheimer}.  


\subsection{Gated filtering technique}

A method that has been discussed and successfully applied for TOF in the past is to periodically cut off the electron flux \cite{bib:macetof}. While there it has been used as a band-pass filter, where all signals with a TOF outside a certain time window have been rejected and a classic, non-integrated beta spectrum has been measured, this technique might as well be applied for TOF spectroscopy. In the case of KATRIN,  periodic filtering could be achieved by a high-frequency modulation of the source or the pre-spectrometer potential. 

\begin{figure}
\centering
\includegraphics[width=0.7\textwidth]{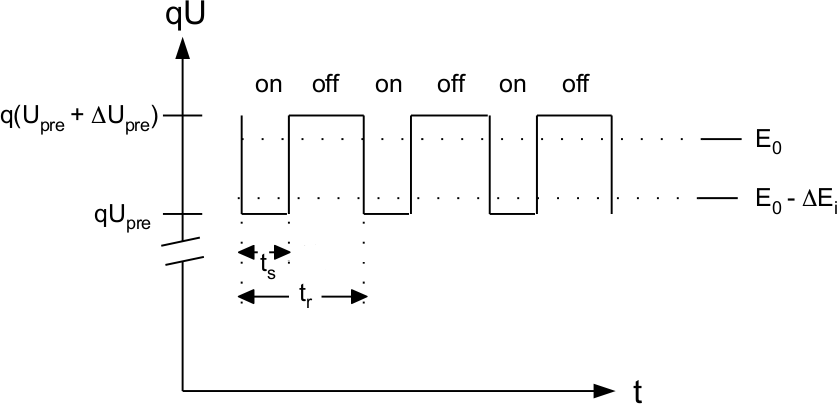}
\caption{Timing parameters of the gated filter. X axis: time. Y axis: pre-spectrometer retarding potential. At the lower filter setting all electrons of the interesting region of width 
$\Delta E_\mathrm{i}$ 
below the endpoint $\Eep$  are transmitted while at the higher setting all electrons are blocked. }
\label{fig:gatedfilter}
\end{figure}

The principle in this case is to switch between two 
settings.  In one setting a pre-spectrometer potential $q(U\sub{pre}+\Delta U\sub{pre}) > \Eep$ is chosen, to block  completely the  flux of $\beta$-electrons. In the other setting the retarding potential of the pre-spectrometer is set to $qU\sub{pre} < E_0 - \Delta E_\mathrm{i}$, leading to the full transmission of all electrons from the interesting energy region $[\Eep-\Delta E_\mathrm{i},\Eep]$ and allowing to 
do TOF spectroscopy (see Fig. \ref{fig:gatedfilter}). The region of interest of width 
$\Delta E_\mathrm{i}$
of a few 10 eV requires moderate switching voltages $\Delta U\sub{pre} \approx -200$ V, taking into account the energy resolution of the pre-spectrometer of about 100~eV. While pulsing the source potential is the traditional approach as in \cite{bib:macetof}, in a dual spectrometer set-up like KATRIN it thus is more convenient to vary the retarding potential of the pre-spectrometer. This has the advantage that, in contrast to pulsing at the source, the potential for the `on'-setting does not need to be precise as long as a transmission of the region of interest is guaranteed. 

\subsubsection{Timing Parameters}

A periodically gated flux can in the simplest case be described by two timing parameters. The first one is the period $t_r$ with which the flux is gated. After each period the detector clock is reset. The second one is the time $t_s$ in which the gate is open in each period. The ratio of $t_s$ and $t_r$ gives the duty cycle

\begin{equation}
\eta = \frac{t_s}{t_r}.
\end{equation}

This method uses no direct measurement of the starting times but restricts them to certain intervals of length $t_s$. That is equivalent to knowing the starting time with certain uncertainty. Thus, for $t_s \longrightarrow 0$ and sufficient period lengths $t_r$, infinitesimally sharp starting times with infinitesimally low luminosity are obtained. If $t_s$ is extended, the luminosity increases and the time uncertainty grows. The uncertainty is given by a uniform probability distribution in the interval $[0; t_s]$. As the measured TOF spectrum is then given by the convolution with the detection time and the starting time distribution,

\begin{equation}
\label{eq_ts}
\left(\tofspec\right)_{t_s} = \tofspec \otimes N(\sigma_d) \otimes U(0, t_s) ,
\end{equation}
where $N(\sigma_d)$ is the Gaussian uncertainty profile of the detection time at the detector 
and $U(0, t_s)$ is the uniform uncertainty due to the gate. If the detector clock is periodically reset with times $t_r$ then some electrons with flight times $> t_r$ might hit the detector in a later period. Therefore, the final measured spectrum is a superposition of all contributing time distributions \eqref{eq_ts}, shifted by multiples of $t_r$ and finally cut off at 0 and $t_r$:

\begin{eqnarray}
\label{eq_tstr}
\left(\tofspec \right)_{t_s, t_r}(t) & = 0 & \qquad t < 0 \non\\
& = \sum _{n=0}^\infty \tofspec_{t_s}(t + n \cdot t_r) & \qquad 0 \leq t \leq t_r \non\\
& = 0 & \qquad t > t_r
\end{eqnarray}

As $> 99.5 \%$ of the flight times lie within $\lesssim \val{50}{\mu s}$\footnote{under the condition that the retarding potential is at least some eV below the endpoint}, all contributions with $n \cdot t_r \gtrsim t_s + \val{50}{\mu s}$ can be neglected. The resulting measured time spectrum \eqref{eq_tstr} is strictly speaking not a spectrum of flight times, but rather of detection times for a periodically reset detector clock.

An illustrative TOF spectrum of simulated measurement data according to \eqref{eq_tstr} is shown in Fig. \ref{fig:gatedfilterexample}. The two characteristics mentioned that describe a simple periodic gate show clear signatures in the curve. The uniform start time distribution within $[0,t_s]$ imposed on the spectrum by the uniform uncertainty \eqref{eq_ts} leads to a clear broadening of the shape. Since the time precision is lower than in the tagger case the broadening is more pronounced. In addition, the smearing with the step function leads to steeper edges than the Gaussian one. A clear sign of the detector resets are the residuals from former gate periods at the beginning of the spectrum. The effects of the timing parameters on the spectral shape allow some preliminary predictions on the effects of the performance:

\begin{figure}
\centering
\includegraphics[width=\linewidth]{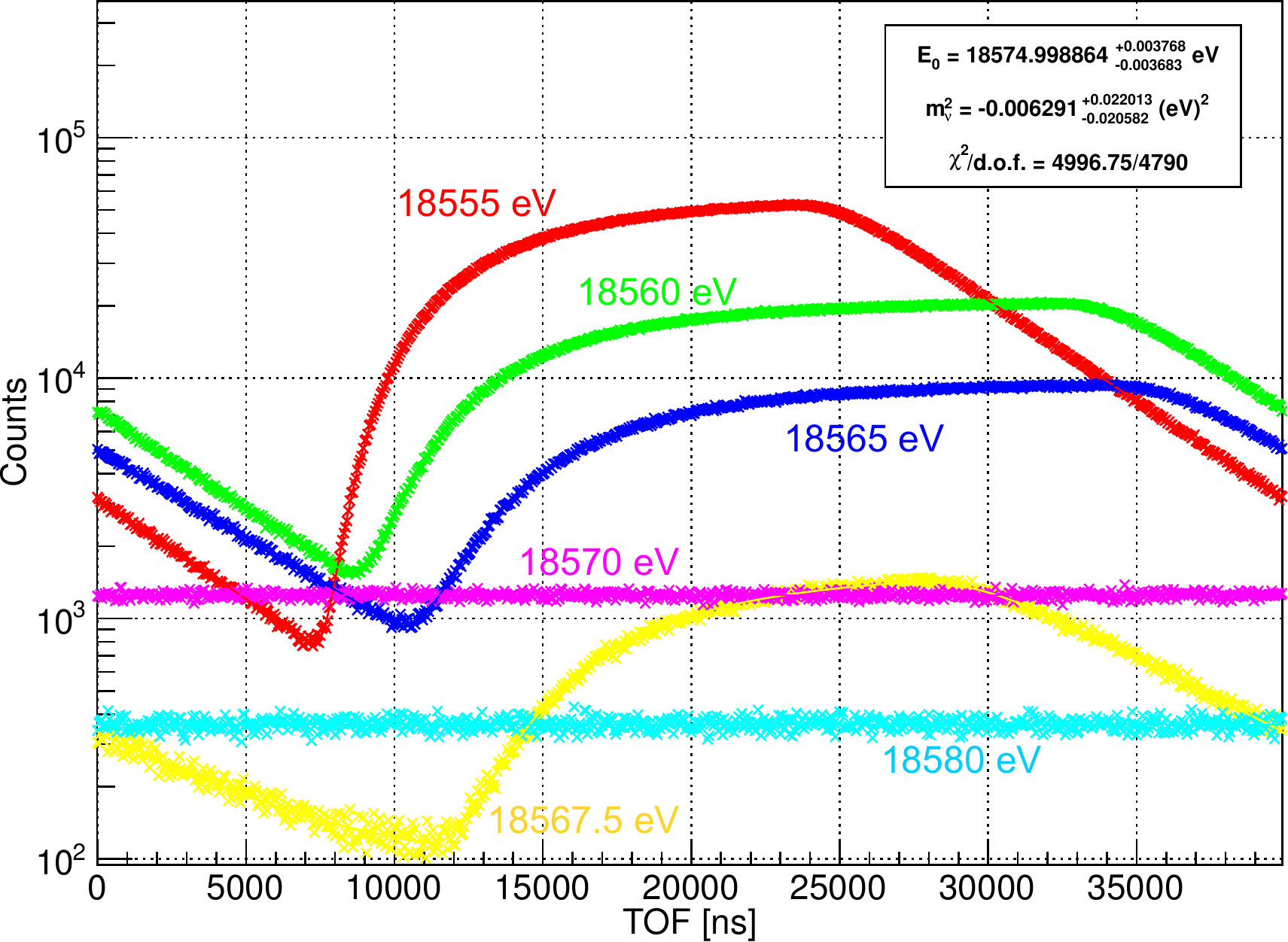}
\caption{Example of simulated data of a TOF spectrum with a gated filter and fit. The colours correspond to the main spectrometer retarding energies. $t_r$ is held constant at $\val{40}{\mu s}$ while $t_s$ and the measurement times per potential step have been chosen to match the optimal distribution stated in table \ref{tb_measoptpulsed}. A total measurement time of three years has been assumed. On the left side of the spectrum the residuals from earlier cycles can be seen which emerge continuously from the end of the spectrum. The peaks exhibit effects of the convolution with the uniform start time distribution. 
For retarding energies of 18570~eV and 18570~eV the gated filter was always open ($t_s=t_r$) yielding
time-independent count numbers.
The other parameters in the simulation were $\bkatrin = 10^{-2}$/s, $E_0 = \val{18575.0}{eV}$ and $\mnusq = 0$.}
\label{fig:gatedfilterexample}
\end{figure}

\begin{itemize}
\item For constant $t_s$, reducing $t_r$ will increase the duty cycle. However, more residuals from former periods contaminate the spectrum. That acts like a non-uniform background. Duty cycle and residual contributions need to be balanced.
\item For constant $t_r$, reducing $t_s$ will reduce the time uncertainty. In contrast, the duty cycle will be reduced, resulting in a lower count-rate. Here, the timing and the duty cycle need to be balanced.
\end{itemize}

Due to the trade-off between timing, duty cycle and residual background, the timing parameters need to be optimized. A global optimization of $t_r$ and $t_s$ has only shown a significant change in $\smunu$ for extreme input values. However, an individual optimization of $t_s$ together with the measurement time contribution for each retarding energy makes sense, as for higher retarding potentials the count-rate drops while the neutrino mass information grows, so a higher duty cycle is needed. As the gated filter is less time-of-flight-sensitive than an ideal TOF measurement, a higher number of retarding energies as well as measurements at lower retarding energies are necessary.

\subsubsection{Results}

\begin{table}
\small
\caption{\label{tb_measoptpulsed}Average statistical uncertainty $<\smunu> ~=~ <\frac{1}{2} (|{\Delta\mnusq}_-| + | {\Delta\mnusq}_+|)>$ (arithmetic mean of positive and negative error of 10 simulations and fits), average fit parameters $<\bar\Eep>$ and  $<\bar\mnusq>$, as well as Pearson's correlation coefficient $R(\Eep,\mnusq)$ for uniform and optimized distribution of a gated filter setup. Assumed are three years measurement time with $\mnusq = \unit[0]{eV^2/c^4}$, $E_0 = \unit[18575.0]{eV}$ and $qU=18555$ eV as lowest retarding energy. The pulse period $t_r$ was held constant at $\val{40}{\mu s}$.}
{\footnotesize
\begin{tabular}{@{}lllllllllll}

\br
\multicolumn{6}{@{}c}{(duty cycle $t_s/t_r$, measurement time fraction) at $qU = $} & $\smunu$ & $<\bar\Eep>$&$<\bar\mnusq>$&$R(\Eep , \mnusq)$\\
18555 V & 18560 V & 18565 V & 18567.5 V & 18570 V & 18575 V &{ [eV$^2$/c$^4$] }&{ [eV$^2$/c$^4$] }& { [eV] }&{ [eV$^2$/c$^4$] }\\
\mr
(0.5, $\frac{1}{6}$) & (0.5, $\frac{1}{6}$) & (0.5, $\frac{1}{6}$) & (0.5, $\frac{1}{6}$) & (0.5, $\frac{1}{6}$) & (0.5, $\frac{1}{6}$) &  \lineup 0.025 & \lineup 18574.9983 & \lineup -0.0056 & 0.9230\\
(0.4, $\frac{1}{13}$) & (0.6, $\frac{1}{13}$) & (0.6, $\frac{2}{13}$) & (0.4, $\frac{1}{13}$) & (1.0, $\frac{4}{13}$) & (1.0, $\frac{4}{13}$) & \lineup 0.021 & \lineup 18575.0006 & \lineup 0.0049 & 0.8914\\
\br

\end{tabular}}
\end{table}
The results of a rough optimization run with 6 retarding potentials, giving 12 free parameters, are shown in Table \ref{tb_measoptpulsed}. The highest retarding energy is above the endpoint that was assumed, thus being sensitive to the background level. Starting with $\eta = 0.5$ and a uniform distribution, each parameter has been scanned successively and set to the position of the local minimum. This has been repeated until the improvements per iteration are sufficiently small. The optimum has been found after 5 iterations. It can be seen that the optimization of duty cycles and measurement times provides an improvement of $\sim$ 30 \% compared to the uniform distribution with $\eta = 0.5$. The obtained result of $\smunu = \unit[0.021]{eV^2/c^4}$ is nearly identical with the standard KATRIN value of $\smunu = \unit[0.020]{eV^2/c^4}$ for the case of 20 eV difference between endpoint and lowest retarding energy. It remains open if a more detailed parameter optimization is able to increase the sensitivity.

\section{Conclusion and outlook}

\begin{figure}
	\centering
	\includegraphics[width=0.7\lw]{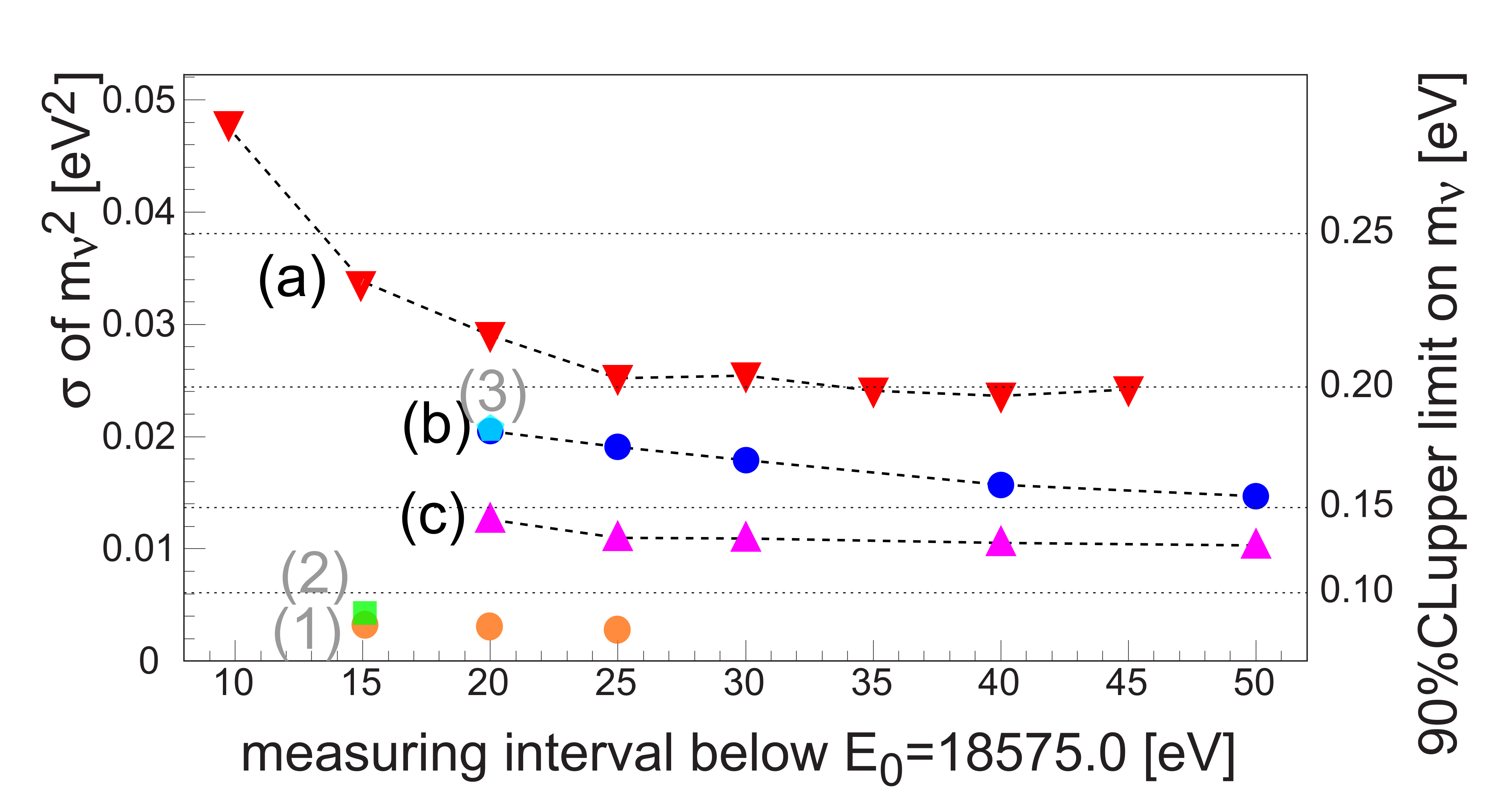}
	\caption{Statistical uncertainty $\smunu$ (3 years measurement time) and corresponding 90 $\%$ C.L. upper limit on $\mnu$ as a function of the analyzed interval for different configurations of standard and TOF mode. Standard mode: (a) uniform measurement time; (b) optimized measurement time; (c) optimized measurement time, but background rate $b$ = 1 mHz instead of 10 mHz as for (a) and (b).
	Results (a)-(c) and figure adapted from \cite{bib:designreport}. TOF spectroscopy (this work): (1) optimized measurement time, no background and infinite time resolution; (2) same as (1) for one examplary measurement interval with a non-zero background rate $b$ = 10 mHz; (3) gated filter with optimized measurement time and optimized duty cycle again for one examplary measurement interval.
Since it is well-known (e.g. \cite{bib:designreport}) that the systematic uncertainties increase with increasing measurement interval below the endpoint $\Eep $ we have concentrated our time-of-flight spectroscopy simulations to short measurement intervals, because otherwise any 
improvement in statistics might be overruled by systematic uncertainties.	}
	\label{fig:statisticalerror}
\end{figure}

A TOF spectroscopy mode could in principle provide significant improvements in the statistical neutrino mass sensitivity compared to a standard MAC-E-Filter mode. The study especially revealed the following information.

\begin{itemize}
\item In the standard mode it is necessary to measure at lower retarding potentials, for instance at KATRIN down to 30 eV below the endpoint, with a large number of measurement points. 
\item Using a TOF mode in contrast, it is sufficient to consider two or more retarding potentials only which may be even more close to the endpoint while improving the statistical uncertainty.
\item This suggests that even the systematic uncertainty can be reduced with a TOF mode as the systematics grow with lower retarding potentials.
\end{itemize}

For a quantitative analysis of the improvement potential of the TOF mode relative to the standard mode one may consider fig. \ref{fig:statisticalerror}, where the statistical uncertainty of $\mnusq$ is plotted as a function of the measurement interval below the endpoint $\Eep$ (difference between lowest retarding potential and the endpoint $\Eep$ using  $\Eep = 18.575$ keV). Compared with the reference value of KATRIN, $\smunu = \unit[0.018]{eV^2/c^4}$ (see figure \ref{fig:statisticalerror} curve (b) for measurement interval of 30 eV), a statistical improvement of up to a factor 5 is possible in the optimal case (fig. \ref{fig:statisticalerror} (1)), equivalent to a factor of more than 2 in statistical sensitivity of $\mnu$. It can be shown (compare the difference in fig. \ref{fig:statisticalerror} between curves (b) and (c) w.r.t. point (2)) that this improvement factor is essentially not caused by neglecting the background but by intrinsic advantages of the method itself. A total improvement factor needs to take the systematics into account, which may only be simulated precisely if the measurement method is sufficiently known. This is especially true since both systematic and statistical uncertainty depend on the choice of retarding potentials, where an optimal trade-off has to be found. 

Considering the measurement method, up to now no  technique has been demonstrated that would  allow a highly 
precise determination of the time-of-flight of the electrons without disturbing their energy significantly. However, there is no fundamental obstacle to a measurement of this kind, and the main difficulty is one of extracting a sufficient and controlled amount of energy from the electron in flight.  The most promising approach, nevertheless very challenging, is detection of the burst of cyclotron radiation as the electron passes through a zone of high magnetic field.  If such a method existed, it would have the advantage of being not only a very sensitive implementation of the TOF mode, but also could significantly suppress backgrounds, depending on the total signal rate.

The method of a periodic gate, which  has been tested in the context of the Mainz experiment \cite{bib:macetof}, may be applied to TOF spectroscopy. The simulations, using a rough parameter optimization, show that its sensitivity is comparable with the standard MAC-E mode (fig. \ref{fig:statisticalerror} point (3) and curve (b)). Hence, whether a TOF mode based on gated filtering is an improvement depends mainly on whether the systematics of that method are better or worse, compared with the standard method.

Apart from the  measurement of the mass of the light active neutrinos, the gated filter may be of special interest for the detection of keV sterile neutrinos. In the warm dark matter (WDM) scenario, these additional neutrino mass states contribute to a large fraction of the dark matter in the universe and are weakly mixed with the electron neutrino \cite{bib:wdm}.  As this mixing would give rise to a subtle 'kink' in the beta spectrum at $E_0 - m_4$, where $m_4 = \mathcal{O}(\mathrm{keV})$ is the additional mass state, it might be useful to utilize a gated filter setup with narrow duty cycles providing sharp timing. The sharp gated filter reduces the statistics but a keV sterile $\nu$ signal will probably be limited by systematics in any case. There is the clear expectation that the TOF method will help in reducing the systematic uncertainty. The reason for this is that the TOF method as a differential method is able to disentangle the more energetic spectral parts, bearing no information about the sterile $\nu$, from the sensitive parts near the potential barrier $qU$, within the region of the 'kink'. Thus, a significant part of the systematics, which would fully contribute in an integrating method, in which only the count-rate is measured, would have no effect on the sterile $\nu$ sensitivity.

For further investigation, a detailed study of systematics of a gated filter driven TOF mode is necessary. This could comprise experimental studies with an angular selective electron gun \cite{bib:valerius2009}, theoretical considerations and Monte Carlo simulations. It may also be useful to investigate the statistics of the gated filter in more detail, considering more parameters to be optimized.

\ack
This work is partly supported by BMBF under contract number 05A11PM2 and by the US Department of Energy under Grant DE-FG02-97ER41020.

\end{document}